\documentclass[journal]{IEEEtran}
\usepackage{amsfonts}
\usepackage{graphicx}
\usepackage{amsfonts,color}
\usepackage[cmex10]{amsmath}
\usepackage{amssymb}
\usepackage{epsfig}
\usepackage[noadjust]{cite}
\usepackage{latexsym}
\usepackage{epstopdf}
\usepackage{amsmath, amsthm, amssymb}
\usepackage{verbatim}
\usepackage{subcaption}
\usepackage[utf8]{inputenc}
\usepackage{algorithmic}
\usepackage{algorithm}
\usepackage{enumitem}
\usepackage{soul}

\definecolor{orange}{RGB}{255,107,0}

\ifCLASSINFOpdf
\else
\fi
\usepackage{amsmath}
\usepackage{algorithmic}
\usepackage{array}

\begin{document}
\title{Fast Algorithms for Joint Multicast Beamforming and Antenna Selection in Massive MIMO}

\author{Mohamed S. Ibrahim,~\IEEEmembership{Student~Member,~IEEE,}
        Aritra~Konar,~\IEEEmembership{Member,~IEEE,}
        and~Nicholas~D.~Sidiropoulos,~\IEEEmembership{Fellow,~IEEE}
\thanks{Mohamed~S.~Ibrahim,~Aritra~Konar,~and~Nicholas~D.~Sidiropoulos are with the Department of Electrical and Computer Engineering, University of Virginia, Charlottesville, VA, 22904 USA (e-mail (mi6cw,aritra,nikos@virginia.edu))}}

\maketitle

\begin{abstract}
Massive MIMO is currently a leading physical layer technology candidate that can dramatically enhance throughput in 5G systems, for both unicast and multicast transmission modalities. As antenna elements are becoming smaller and cheaper in the mmW range compared to radio frequency (RF) chains, it is crucial to perform antenna selection at the transmitter, such that the available RF chains are switched to an appropriate subset of antennas. This paper considers the joint problem of multicast beamforming and antenna selection for a single multicast group in massive MIMO systems. The prior state-of-art for this problem relies on semi-definite relaxation (SDR), which cannot scale up to the massive MIMO regime. A successive convex approximation (SCA) based approach is proposed to tackle max-min fair joint multicast beamforming and antenna selection. The key idea of SCA is to successively approximate the non-convex problem by a class of non-smooth, convex optimization problems. Two fast and memory efficient first-order methods are proposed to solve each SCA subproblem. Simulations demonstrate that the proposed algorithms outperform the existing state-of-art approach in terms of solution quality and run time, in both traditional and especially in massive MIMO settings.
\end{abstract}

\IEEEpeerreviewmaketitle

\section{Introduction}\label{intro}
\IEEEPARstart{O}{wing} to the tremendous growth of Internet-connected devices through wireless networks, collectively referred to as the Internet of Things (IoT), providing high throughput to such devices constitutes one of the major challenges for fifth generation (5G) wireless systems~\cite{osseiran2014scenarios,mobile2011global}. Massive MIMO, which refers to equipping the base station (BS) with hundreds of transmit antennas, is considered to be one of the key physical layer techniques that can support the expected increase in data rates ~\cite{larsson2014massive,boccardi2014five}. Theoretical studies have demonstrated that massive MIMO can result in a dramatic increase in both spectral efficiency and energy efficiency~\cite{ngo2013energy}, albeit at the expense of high hardware cost and complexity. 

\textit{Multicast beamforming} is a simple but powerful technique that exploits channel state information at the BS to create a multi-spot beam pattern that can be used to deliver high throughput {\em common} information to a set of users or devices \cite{sidiropoulos2006transmit,lecompte2012evolved}.  While multicast beamforming was originally developed for video (e.g., event) streaming to mobile users, emergent IoT and vehicular communication applications can benefit from multicast beamforming for a range of services, from road awareness to software updates.

A key obstacle in implementing large scale antenna systems is the cost and complexity of the associated hardware. Owing to the cost, size and power consumption of radio frequency chains, devoting an RF chain to each antenna element in massive MIMO scenarios is not feasible. This is because each RF chain includes a mixer, a power amplifier, and analog-to-digital (A/D) and digital-to-analog (D/A) converters. Since, from the practical point of view, the number of available RF chains will be less than the number of transmit antennas, we need to switch the available RF chains to the most appropriate subset of antennas.

Determining the latter is a combinatorial problem, which is compounded by the fact that assessing the multicast capacity of any given subset is a hard problem in its own right \cite{sidiropoulos2006transmit}. This renders the problem of designing capacity- and computation-efficient algorithms for jointly selecting a subset of antennas and finding the associated beamforming vector of paramount importance. This is precisely the problem we tackle in this paper. 
           
\subsection{Prior art and Motivation}
Single-group max-min fair multicast beamforming under a sum power constraint was first formulated and shown to be NP-hard in~\cite{sidiropoulos2006transmit}. The formulation was later extended to multiple co-channel multicast groups in~\cite{karipidis2008quality}. These papers considered only the traditional multicasting scenario where the number of users is much larger than the number of antennas.  On the other hand, massive MIMO multicasting in cellular networks was first considered in~\cite{xiang2014massive}, where an asymptotic analysis for the optimal beamforming vector at each base station was provided. Then, the work of~\cite{xiang2014massive} was later extended to the multi-group multicast in~\cite{zhou2015joint}, and to the multi-cell multi-group in~\cite{sadeghi2015multi}. 

Since, for massive MIMO scenarios, it is more practical to use per antenna power constraints (PAPCs),~\cite{yu2007transmitter}, than the sum power constraint, a lot of work has considered multicasting with PAPCs. In~\cite{christopoulos2014multicast}, the authors considered the multi-group multi-casting problem under PAPC, where an equally fair multicast multi-group solution using a semi-definite relaxation approach was presented. Then, the authors extended their work to the weighted max-min fair multi-group multicast beamforming~\cite{christopoulos2014weighted}. In~\cite{christopoulos2015multicast}, the authors used a successive convex approximation (SCA) approach~\cite{marks1978general,razaviyayn2013unified,beck2010sequential} to tackle the max-min fair multi-group multicasting problem under PAPC, where each SCA sub-problem was formulated as a second order cone program. A subsequent work~\cite{7874154} used an alternative reformulation which
entails solving a sequence of per-antenna minimization problems within a bisection framework. Then, an ADMM
algorithm is proposed for (approximately) solving these per-antenna minimization problems.
Later on, \cite{konar2017fast} considered the max-min fair single-group multicasting problem, where a SCA framework was adopted to develop low-complexity first order based methods. Then a follow-up work proposed a sub-gradient method to solve the max-min fair single group multicasting with sum power constraint~\cite{8446006}. In all of the aforementioned papers, antenna selection has not been considered.

Antenna subset selection for point-to-point MIMO links has been considered in the early 2000's, see \cite{sanayei2004antenna} and references therein, and \cite{nai2010beampattern}. For receive beamforming, antenna selection was considered in \cite{dua2006receive}, and for downlink capacity maximization of massive MIMO systems in~\cite{gao2015massive}. The latter carried out measurements in a multi-user MIMO system at $ 2.6 $ GHz, using a linear and a cylindrical array (each equipped with $ 128 $ elements) at the base-station and reported large power variations across the antenna elements in both line-of-sight and non line-of-sight scenarios. The authors concluded from their experiments that these power variations makes some antennas more ``useful'' than others, and proposed to use antenna selection to select the most appropriate antennas. This suggests that antenna selection has much potential as means of achieving a favorable performance-complexity tradeoff in Massive MIMO systems.

Due to the combinatorial nature of antenna selection problems, various approximation algorithms have been proposed, ranging from semi-definite relaxation (SDR)~\cite{mehanna2013joint}, greedy algorithms~\cite{konargreed}, and machine learning algorithms~\cite{ibrahim2018learning,joung2016machine}. For the traditional multicast scenario, joint max-min fair multicast beamforming and antenna selection was initially studied in~\cite{park2008capacity} for systems with relatively few transmit antennas. Follow-up work in \cite{mehanna2013joint} proposed a SDR-based approach that can handle the antenna selection and beamforming design tasks jointly without the need to consider all subsets. Although the SDR-based approach has been shown to be effective in identifying high-quality, approximate solutions for the joint problem~\cite{mehanna2013joint}, it lifts the optimization task into an equivalent higher dimensional problem with a rank constraint. Then, the problem is relaxed by dropping the rank constraint. Since the relaxed problem's solution is not rank one in general, an extra randomization step, which adds to the computational burden of the overall algorithm, is required to obtain an approximate solution. Hence, this approach suffers from high computational and memory complexity in massive MIMO settings. This motivates the pursuit of fast, low-complexity, efficient algorithms that can provide high quality approximate solutions and are scalable to massive MIMO scenarios.

\subsection{Contribution}
The goal of this paper is to solve the problem of joint multicast beamforming and antenna selection in a single-group massive MIMO setting with PAPCs.  A high performance, low complexity successive convex approximation (SCA) based approach is proposed for solving the joint problem. Although SCA has not been considered before for joint multicast beamforming and antenna selection, our SCA framework compared to that of~\cite{christopoulos2015multicast,7874154}- which studied the max-min fair multicasting without antenna selection- features the following;  (i) it approximates the objective function of the problem directly without introducing any additional variables or constraints, (ii) exploits the structure inherent in each SCA subproblem by using specially tailored low-complexity first-order methods, and (iii) guarantees convergence to a stationary point.

As the objective function of our problem is the sum of a convex and a non-convex function, the proposed SCA approach seeks to successively majorize the non-convex part of the objective. After replacing the non-convex function with a convex surrogate, we are left with a non-smooth, convex optimization subproblem at each iteration. Then, the solution of each SCA subproblem is used as the point about which the convex surrogate function will be constructed in the next iteration.

Since the overall complexity of the SCA approach is determined by the cost of solving each subproblem, this paper proposes two fast, first-order based methods (FOMs) to handle each SCA subproblem. First, each subproblem is reformulated in a form for which the Consensus Alternating Direction Method of Multipliers (C-ADMM)~\cite{boyd2011distributed} can be applied. In fact, C-ADMM exhibits several advantages; it is well suited for parallel and distributed optimization by carefully splitting the objective to a sum of separate functions. Moreover, C-ADMM yields very efficient updates (sometimes in a closed form). Finally, for convex problems, which is our case, it guarantees convergence for any constant positive penalty parameter.  

Second, the saddle-point mirror-prox~(SP-MP)~\cite{nesterov1983method} algorithm is proposed to solve each SCA subproblem. It is shown how group sparsity and the dual norm ~\cite{francis} can be exploited to reformulate each SCA subproblem as a maximization of a bilinear function over a convex set. 
This special first-order based algorithm can be used to efficiently solve each SCA subproblem, featuring low per iteration complexity which is linear in the number of variables. This renders the algorithm very efficient in massive MIMO scenarios.

In order to demonstrate the superior performance of the proposed algorithms, two settings for the joint problem are considered. First, the traditional multicasting scenario under a sum power constraint, where the number of users is much larger than the number of antennas at the transmitter. In such a scenario, the proposed low-complexity algorithms are compared with the state-of-art SDP-based approach presented in~\cite{mehanna2013joint}. Simulations demonstrate that the proposed algorithms provide comparable quality solutions to the SDP-based one, but at far lower complexity. Second, a massive MIMO setting is considered under PAPCs. This is the first attempt to design low-complexity algorithms that can efficiently approximate the joint problem in this setting. Simulations reveal that the proposed FOM-based SCA algorithms strike the right balance between performance and complexity for this regime. 

An early version of part of this work has appeared in~\cite{ibrahim2018prox}, which proposed the SP-MP algorithm to solve the max-min fair problem with antenna selection for the traditional multicasting scenario. This journal version focuses on the massive MIMO scenario and brings in the C-ADMM algorithm to solve each SCA subproblem, along with a  comprehensive suite of simulation experiments to demonstrate the efficacy of the proposed algorithms in massive MIMO settings.  
                    
          \subsection{Organization}
The remainder of this paper is organized as follows. Section~\ref{Pre} describes the system model and introduces some definitions. Section~\ref{Form} presents the problem formulation. The proposed SCA procedures are discussed in~Section~\ref{SCA}, while the proposed algorithms are presented in~Section~\ref{prop_alg}. Then, simulations results are provided in Section~\ref{SR}. Finally, conclusions are drawn in Section~\ref{Conc}.   

          \subsection{Notation}
We use upper and lower case bold letters to denote
matrices and column vectors, respectively. We use $(.)^T$,
$(.)^H$ to  denote the transpose and Hermitian (conjugate) transpose, respectively.
Scalars are represented in the normal face, while calligraphic font is used to denote sets. ${\rm I\!Re}\{\}$ and $ {\rm I\!Im}\{\}  $ denote the real and imaginary operators, respectively. $ \lVert.\rVert_2 $ and $ \lVert.\rVert_1 $ denote the $ \ell_2 $- and $ \ell_1 $-norms, respectively. If $ f() $ is differentiable, its gradient operator is denoted by $ \nabla f(.) $. $ \mathbb{R}^N $ and $ \mathbb{C}^N $ represent the $N$-dimensional real and complex Euclidean spaces, respectively. Finally, $ {\bf I}_N $ and $ {\bf 0}_N $ denote the $ N \times N $ identity matrix and the $ N \times 1 $ zero vector, respectively.
 

\section{Preliminaries}\label{Pre}
\subsection{Basic Model}
Consider a wireless scenario consisting of a single BS equipped with $ N $ antennas serving $ M $ single-antenna users in the downlink.
Let $ {{\bf h}_m} $ denote the $ N \times 1 $ complex vector that models the propagation loss and frequency-flat quasi-static downlink channel from the BS to user $ m \in \mathcal{M}:=\{1, \cdots, M\} $. The BS utilizes a beamforming vector $ {\bf w} \in \mathbb{C}^{N} $  to transmit a common information bearing signal $s \in \mathbb{C}$ to the $ M $ users. The corresponding received signal at the $ m $-th user is given by
\begin{equation}
y_m  = {\bf h}_m^H {\bf w} s + n_m, \forall \; m \in \mathcal{M}
\end{equation}                    
\noindent where $ n_m $ is zero-mean Gaussian noise at the $ m $-th user with variance $ \sigma_m^2 $, and is assumed to be independent of $ s $ and $ {\bf h}_m $. It is further assumed that the channel vectors $ {\bf h}_m $ and their respective $ \sigma_m^2 $, $ \forall \; m \in \mathcal{M} $, are known at the BS. The received SNR at the $ m $-th user can be expressed as
\begin{equation}
\frac{\lvert \mathbf{h}_m^H\mathbf{w} \rvert^2 }{\sigma^2_m}  = \mathbf{w}^H \mathbf{A}_m \mathbf{w},
\end{equation}
where $ \mathbf{A}_m := \frac{\mathbf{h}_m \mathbf{h}^H_m }{\sigma^2_m} \succeq \mathbf{0}$, $ \forall \; m \in \mathcal{M} $. Since the common message needs to be decoded by all receivers, the maximum achievable common rate is determined by the ``bottleneck'' user, i.e., the user with minimum received SNR. Therefore, the design problem at the BS is to maximize the minimum received SNR subject to sum power constraints \cite{sidiropoulos2006transmit}, which can be formulated as
\begin{subequations}\label{BM1}
	\begin{align}
	& \underset{{\bf w}\in \mathbb{C}^N}{\max} ~ \underset{m \in \mathcal{M}}{\min} ~ \mathbf{w}^H \mathbf{A}_m \mathbf{w}\\
	& \text{s.t.} \quad \ \| {\bf w} \|_2^2 \leq P
	\end{align}
\end{subequations}
where $ P $ represents the available power at the BS. Problem (\ref{BM1}) is non-convex, as the point-wise minimum of convex quadratics is a non convex-function. In addition, the problem has been proven to be NP-hard~\cite{sidiropoulos2006transmit}, when $ N \leq M $. An equivalent problem seeks to minimize the transmitted power subject to per user quality-of-service (QoS) constraint~\cite{sidiropoulos2006transmit}. Problem (\ref{BM1}) has been tackled by several algorithms in the literature. For instance, in~\cite{sidiropoulos2006transmit}, a semidefinite relaxation approach was adopted for computing a high-quality approximate solution. Moreover, iterative procedures such as alternating minimization~\cite{demir2014alternating}, and successive convex approximation, ~\cite{tran2014conic,bornhorst2011iterative}, have been applied to obtain approximate solutions for the QoS version of (\ref{BM1}). Note that the SCA approach outperforms all other algorithms in terms of solution quality and complexity, when $ N \leq M $.

On the other hand,~\cite{konar2017fast} has studied the max-min fair multicast beamforming problem, when $ N > M $. This case is applicable in massive MIMO scenarios,~\cite{rusek2013scaling,lu2014overview}, where the base station is equipped with a large number of antennas. Note that it is more practical to consider per antenna power constraints instead of a sum power constraint (\ref{BM1}b) when the antennas are fed by different power amplifiers. The per-antenna power constrained problem is  
 \begin{subequations}\label{BM2}
 	\begin{align}
 	& \underset{{\bf w}\in \mathbb{C}^N}{\max} ~ \underset{m \in \mathcal{M}}{\min} ~ \mathbf{w}^H \mathbf{A}_m \mathbf{w}\\
 	& \text{s.t.} \quad \ | {\bf w}(i) |^2 \leq P_i, \quad \forall \; i \in \{1,\cdots,N\}
 	\end{align}
 \end{subequations} 
which is still non-convex. In order to approximate (\ref{BM2}), the authors of ~\cite{konar2017fast} proposed a SCA-approach using specialized FOMs to solve each subproblem. For large $N$, this approach in \cite{konar2017fast} demonstrated considerable computational savings compared to SDR and SCA using generic convex programming solvers for solving each subproblem.  

\subsection{Antenna Selection}
Let $ K < N $ denote the number of RF chains available at the BS. Thus, only a subset of $ K $ antennas can be connected to the available RF chains using analog switches. The goal here is to jointly select the ``best'' $ K $ out of $ N $ antennas and design the corresponding beamforming vector associated with the selected antennas, such that the minimum received SNR is maximized subject to PAPCs. The resultant problem can be posed as
\begin{subequations}\label{AS-1}
	\begin{align}
	& \underset{{\bf w}\in \mathbb{C}^N}{\max} && \hspace{-1cm} \underset{m \in \mathcal{M}}{\min} ~ \mathbf{w}^H \mathbf{A}_m \mathbf{w}\\
	& \text{s.t.}   && \hspace{-1cm} |{\bf w}(i) |^2 \leq P_i, \forall \;i \in \{1,\cdots,N\}, \\
	&&&  \hspace{-1cm} \|{\bf w}\|_0 \leq K \label{eq:subn5c}
	\end{align} 
\end{subequations}
The cardinality constraint \eqref{eq:subn5c} is non-convex, \emph{and} non-differentiable, which renders the problem more computationally challenging. One possible approach is to apply exhaustive search over the $ \binom{N}{K} $ possible antenna subsets, followed by solving a reduced problem of the form  (\ref{BM2}), using the algorithm in \cite{konar2017fast}, for each subset. However, the computational complexity of exhaustive search becomes prohibitive even for modest values of $ N $, which renders this approach impractical.    

Instead of imposing the cardinality constraint, a reasonable alternative is to augment the objective function of \eqref{AS-1} with a cardinality penalty in order to promote sparsity in $\bf w$. Thereafter, the cardinality function can be replaced by the $ \ell_1 $-convex surrogate to obtain the following problem
\begin{subequations}\label{AS-2}
	\begin{align}
	& \underset{{\bf w}\in \mathbb{C}^N}{\max}~\underset{m \in \mathcal{M}}{\min} ~ \mathbf{w}^H \mathbf{A}_m \mathbf{w} - \lambda \| {\bf w} \|_1\\
	& \text{s.t.} \quad | {\bf w}(i) |^2 \leq P_i, \quad \forall \; i \in \{1,\cdots,N\}
	\end{align}
\end{subequations}    
where $ \lambda > 0 $ is a tuning parameter that controls the sparsity of solution, i.e., increasing $ \lambda $ yields a more sparse solution. Therefore, the value of $ \lambda $ can be tuned until the desired sparsity is attained. Problem (\ref{AS-2}) strikes a balance between maximizing the minimum received SNR and minimizing the number of selected antennas. 


In prior work \cite{mehanna2013joint}, problem (\ref{AS-2}) has been tackled using a SDR-based approach, which has been reported to attain a near optimal solution when $ N \leq M $; albeit at the high cost of having to solve a computationally intensive semidefinite programming problem. This constitutes a key drawback of this approach, since (\ref{AS-2}) needs to be solved many times with different choices of $\lambda$ until the desired sparse solution is obtained. Consequently, using SDR is inefficient for large number of antennas (order of hundreds) at the BS. This motivates the pursuit of computationally efficient algorithms that can perform joint multicasting and antenna selection for massive MIMO scenarios. 

\subsection{Group Sparsity Inducing norms}          
The $ \ell_1 $-penalty (defined as $ \| {\bf w} \|_1 = \sum\limits_{n=1}^{N} |{\bf w}(n)|$) is known to induce sparsity by compelling some entries of the optimal solution $ {\bf w}^{\star} $, depending on the magnitude of $ \lambda $, to be equal to zero. Although it is a weaker measure of sparsity relative to the non-convex cardinality function, it remains the closest convex approximation~\cite{tibshirani1996regression}.  However, as the proposed approach operates in the real domain (see next section), it is inappropriate to use the $ \ell_1 $-norm as a regularizer. This is because switching off an antenna requires setting both its corresponding real and imaginary entries in the beamforming vector to be simultaneously zero. It can be easily seen that the $ \ell_1 $-norm does not guarantee such a requirement. Hence, we propose to utilize an appropriate structured sparsity-inducing norm that can enforce both entries to be zero simultaneously.  

Although there exist many such norms, a popular choice is the group-sparsity promoting mixed $ \ell_{1,2} $-norm, first introduced in~\cite{yuan2006model}. Such a norm takes the form
\begin{equation}\label{GS1}
\|{\bf w} \|_{1,2} :=  \sum\limits_{j = 1}^{N}   \|{\bf w}_j\|_2 
\end{equation} 
where each sub-vector $ {\bf w}_j $ is composed of a group of entries selected from the original vector $ \bf w $. 
Specifically, $ {\bf w}_j = [{\rm I\!Re}\{{\bf w}(j)\},{\rm I\!Im}\{{\bf w}(j)\}]^T$, for $ j \in [N] $. 
Note that the $ \ell_{1,2} $-norm encourages each sub-vector $ \|{\bf w}_g\|_2 $ to be set to zero. Moreover, when each group contains a single entry, the $ \ell_{1,2} $-norm boils down to the $ \ell_1 $-norm. Henceforth, the $ \ell_1 $-norm will be replaced by $ \ell_{1,2} $-norm, and consequently, (\ref{AS-2}) will be replaced by
\begin{subequations}\label{GS2}
	\begin{align}
	& \underset{{\bf w}\in \mathbb{C}^N}{\max}~\underset{m \in \mathcal{M}}{\min} ~ \mathbf{w}^H \mathbf{A}_m \mathbf{w} - \lambda \| {\bf w} \|_{1,2}\\
	& \text{s. to} \quad | {\bf w}(i) |^2 \leq P_i, \quad \forall \; i \in \{1,\cdots,N\}
	\end{align}
\end{subequations} 
On the other hand, the dual norm is important to study sparsity-inducing regularization~\cite{NIPS2008_3418,jenatton2011structured}. The dual norm $ \|.\|^{\star} $ of the norm $ \|.\| $ is defined for any vector $ {\bf x} \in \mathbb{R}^N$ by
\begin{equation} \label{GS3}
\| {\bf x} \|^{\star} = \underset{\|{\bf u}\| \leq 1}{\max} {\bf u}^T{\bf x}
\end{equation} 
The definitions of dual norm and group-sparsity will be utilized to reformulate problem (\ref{GS2}) such that it can be successively approximated, and then, using FOMs to solve each sub-problem. 

\section{Problem Formulation}\label{Form}
Our goal is to develop a low complexity SCA algorithm which can efficiently yield a high-quality approximate solution for (\ref{GS2}). First, let us write (\ref{GS2}) in the real domain by defining the variables $ \bar{{\bf w}} := [{\bf w}_r^T,{\bf w}_i^T]^T  \in \mathbb{R}^{2N}$, where $ {\bf w}_r = {\rm I\!Re}\{{\bf w}\} $ and $ {\bf w}_i = {\rm I\!Im}\{{\bf w}\}  $ represent the real and imaginary components of the complex beamforming vector $ {\bf w} $, respectively. We also define a matrix $ \tilde{{\bf A}}_m \in \mathbb{R}^{2N \times 2N}$ as \\
\begin{equation}
\tilde{{\bf A}}_m := 
\begin{bmatrix} 		
{\rm I\!Re}\{{\bf A}_m\} & -{\rm I\!Im}\{{\bf A}_m\} \\
{\rm I\!Im}\{{\bf A}_m\} & {\rm I\!Re}\{{\bf A}_m\} 
\end{bmatrix} , \forall \; m \in \mathcal{M}
\end{equation}
It can be easily seen that $ {\bf A}_m \succeq {\bf 0} $ iff $ \tilde{{\bf A}}_m \succeq {\bf 0}$. Now, (\ref{GS2}) can be reformulated in terms of real variables as 
\begin{subequations}\label{PF1}
	\begin{align}
	& \underset{\bar{{\bf w}}\in \mathbb{R}^{2N}}{\max} ~ \underset{m \in \mathcal{M}}{\min} ~ \bar{{\bf w}}^T \tilde{{\bf A}}_m \bar{{\bf w}} - \lambda\| \bar{{\bf w}} \|_{1,2}\\
	& \text{s.t.} \quad \| \bar{{\bf w}}(i),\bar{{\bf w}}(i+N) \|^2_2 \leq P_i, \quad \forall \; i \in \{1,\cdots,N\}
	\end{align}
\end{subequations}
Since the objective function of (\ref{PF1}) can be equivalently expressed as
\begin{equation}
 \underset{{\bf w} \in \mathcal{W}}{\max} \;\underset{m \in \mathcal{M}}{\min} f_m({\bf w}) \Leftrightarrow \underset{{\bf w} \in \mathcal{W}}{\min} \; \underset{m \in \mathcal{M}}{\max} -f_m({\bf w}), 
\end{equation}
upon introducing the definition 
\begin{equation}
\bar{{\bf A}}_m := - \tilde{{\bf A}}_m ,  \forall\; m \in \mathcal{M},
\end{equation}
problem (\ref{PF1}) can be equivalently expressed as 
\begin{subequations}\label{PF2}
	\begin{align}
	& \underset{\bar{{\bf w}}\in \mathbb{R}^{2N}}{\min} ~ \underset{m \in \mathcal{M}}{\max} ~ \bar{{\bf w}}^T \bar{{\bf A}}_m \bar{{\bf w}} + \lambda\| \bar{{\bf w}} \|_{1,2}\\
	& \text{s.t.} \quad \| \bar{{\bf w}}(i),\bar{{\bf w}}(i+N) \|^2_2 \leq P_i, \quad \forall \;i \in \{1,\cdots,N\}
	\end{align}
\end{subequations}
By defining the functions
\begin{subequations}
	\begin{align}
	f_1(\bar{{\bf w}}) &:= \underset{m \in \mathcal{M}}{\max} ~ \bar{{\bf w}}^T \bar{{\bf A}}_m \bar{{\bf w}} \\
	f_2(\bar{{\bf w}}) &:=  \lambda\| \bar{{\bf w}} \|_{1,2}
	\end{align}
\end{subequations}
 and the set 
 \begin{equation}
 \mathcal{P} := \{  \bar{{\bf w}} \in \mathbb{R}^{2N}: \| [\bar{{\bf w}}(i),\bar{{\bf w}}(i+N)] \|^2_2 \leq P_i, \forall \; i \in \{1,\cdots, N \}\},
 \end{equation}
 problem (\ref{PF2}) can be compactly expressed as 
\begin{subequations}\label{PF3}
	\begin{align}
	& \underset{\bar{{\bf w}}\in \mathbb{R}^{2N}}{\min} ~  f_1(\bar{{\bf w}}) +  f_2(\bar{{\bf w}})\\
	& \text{s.t.} \quad \bar{{\bf w}} \in \mathcal{P}
	\end{align}
\end{subequations}
 Note that the non-convexity of (\ref{PF3}) is due to $f_1(\bar{{\bf w}})$ alone; the function $f_2(\bar{{\bf w}})$ and the constraint set $\mathcal{P}$ are convex. A low-complexity SCA framework for (\ref{PF3}) will be presented in the next section.

\section{Successive Convex Approximation}\label{SCA}
 The basic idea of SCA is to iteratively solve a sequence of convex problems obtained by locally approximating the non-convex problem (\ref{PF3}) about the current iterate. Starting from an initial feasible point $ \bar{{\bf w}}^{(0)} \in \mathcal{P} $, the approach proceeds as follows. At each iteration $ t \geq 0 $, a locally tight convex upper bound of $f_1(\bar{{\bf w}})$ is constructed about the current iterate $ \bar{{\bf w}}^{(t)} $ in the following manner. First, we write
\begin{equation}
f_1(\bar{{\bf w}}) = \underset{m \in \mathcal{M}}{\max} ~ g_m(\bar{{\bf w}}) ,
\end{equation}
 where $ g_m(\bar{{\bf w}}) := \bar{{\bf w}}^T \bar{{\bf A}}_m \bar{{\bf w}}, \forall \;m \in \mathcal{M}$. Since $ \bar{{\bf A}}_m \preceq 0$, each $g_m(\bar{{\bf w}})$ is a concave function. Hence, local linearization of the function $g_m(\bar{{\bf w}})$ about the point $ \bar{{\bf w}} = \bar{{\bf w}}^{(t)} $ yields the following affine upper bound 
\begin{equation}\label{SCA1}
	\begin{aligned}
	   g_m(\bar{{\bf w}}) &\leq g_m(\bar{{\bf w}}^{(t)}) + \nabla g_m(\bar{{\bf w}}^{(t)})^T(\bar{{\bf w}} - \bar{{\bf w}}^{(t)}) \\
	& = 2(\bar{{\bf A}}_m\bar{{\bf w}}^{(t)})^T\bar{{\bf w}} - \bar{{\bf w}}^{(t)}\bar{{\bf A}}_m\bar{{\bf w}}^{(t)} \\
	& = \bar{{\bf q}}^{(t)T}_m\bar{{\bf w}}+ \bar{b}_m^{(t)}
	\end{aligned}
\end{equation}
where the vector $\bar{{\bf q}}^{(t)}_m := 2\bar{{\bf A}}_m\bar{{\bf w}}^{(t)} \in \mathbb{R}^{2N}$ and the scalar $ \bar{b}_m^{(t)} =  -\bar{{\bf w}}^{(t)}\bar{{\bf A}}_m\bar{{\bf w}}^{(t)}, \forall\; m \in \mathcal{M}$. 
Now, let us define the function
\begin{equation}
 u(\bar{{\bf w}},\bar{{\bf w}}^{(t)}) := \underset{m \in \mathcal{M}}{\max}~\bar{{\bf q}}^{(t)T}_m\bar{{\bf w}}+  \bar{b}_m^{(t)}
\end{equation}
It follows that $ u(\bar{{\bf w}},\bar{{\bf w}}^{(t)}) $ possesses the following properties: 
\begin{enumerate}
	\item[(A1)] $u(\bar{{\bf w}},\bar{{\bf w}}^{(t)})$ is convex in $ \bar{{\bf w}} $, as the maximum of piece-wise affine functions is convex.
	\item[(A2)] $u(\bar{{\bf w}},\bar{{\bf w}}^{(t)})$ is continuous in $ (\bar{{\bf w}},\bar{{\bf w}}^{(t)}) $, but non-differentiable in $ \bar{{\bf w}}$
	\item[(A3)] $u(\bar{{\bf w}},\bar{{\bf w}}^{(t)}) \geq f_1(\bar{{\bf w}}),~\forall~ \bar{{\bf w}},\bar{{\bf w}}^{(t)} \in \mathcal{P}$ 
	\item[(A4)] $u(\bar{{\bf w}}^{(t)},\bar{{\bf w}}^{(t)}) = f_1(\bar{{\bf w}}^{(t)}),~\forall~ \bar{{\bf w}}^{(t)} \in \mathcal{P}$
\end{enumerate}
The last two properties imply that  $u(\bar{{\bf w}},\bar{{\bf w}}^{(t)})$ represents a convex upper bound on the function $ f_1(\bar{{\bf w}}) $, with equality at $ \bar{{\bf w}} = \bar{{\bf w}}^{(t)} $. On replacing $ f_1(\bar{{\bf w}})  $ by its piece-wise linear approximation $ u(\bar{{\bf w}},\bar{{\bf w}}^{(t)}) $ at each iteration $ t $, we obtain the following non-smooth, convex subproblem  
\begin{subequations}\label{SCA2}
	\begin{align}
	& \underset{\bar{{\bf w}}\in \mathbb{R}^{2N}}{\min} ~   u(\bar{{\bf w}},\bar{{\bf w}}^{(t)}) +  f_2(\bar{{\bf w}})\\
	& \text{s.t.} \quad \bar{{\bf w}} \in \mathcal{P}
	\end{align}
\end{subequations}
Iteratively solving such a sequence of convex subproblems yields the following algorithm  
\begin{algorithm} \label{Alg1}
	\caption{SCA}
	\textbf{initialization}: Choose a random starting point $ \bar{{\bf w}}^{(0)} \in \mathcal{P}$ and set $ t := 0 $ \\ 
	\textbf{Repeat:}
	\begin{itemize}
		\item Compute $ \bar{{\bf w}}^{(t+1)} \in \arg\underset{\bar{{\bf w}} \in \mathcal{P}}{\min}~u(\bar{{\bf w}},\bar{{\bf w}}^{(t)}) + f_2(\bar{{\bf w}})$
		\item set $t = t+1$
	\end{itemize}
	\textbf{Until} termination criterion is met
\end{algorithm} 

Note that by construction, the following chain of inequalities holds
\begin{equation}
\begin{aligned}
f_1(\bar{{\bf w}}^{(t+1)}) + f_2(\bar{{\bf w}}^{(t+1)})
&\leq u(\bar{{\bf w}}^{(t+1)},\bar{{\bf w}}^{(t)}) + f_2(\bar{{\bf w}}^{(t+1)})\\
& \leq u(\bar{{\bf w}}^{(t)},\bar{{\bf w}}^{(t)}) + f_2(\bar{{\bf w}}^{(t)})\\
&= f_1(\bar{{\bf w}}^{(t)}) + f_2(\bar{{\bf w}}^{(t)})
\end{aligned}
\end{equation}
where the first inequality is due to (A3), the second inequality holds as $\bar{{\bf w}}^{(t+1)}$ is an optimal solution of (\ref{SCA2}) at the $t^{th}$ iteration, and the final equality is due to (A4). Consequently, the iterate sequence generated by Algorithm $1$ features monotonically non-increasing cost. Additionally, it can be shown that the iterate sequence converges to the set of stationary points of the non-convex problem (\ref{PF3}). The proposition has already been proven in \cite{konar2017fast} for $f_2(\bar{{\bf w}}) = 0$ (i.e., for the case without antenna selection), and relies on establishing a link between the ``first-order'' properties of the non-convex cost function $f_1(\bar{{\bf w}})$ and its convex surrogate $u(\bar{{\bf w}},\bar{{\bf w}}^{(t)})$ at the point $\bar{{\bf w}} = \bar{{\bf w}}^{(t)}$. In the present case, $f_2(\bar{{\bf w}})$ is a non-smooth group-sparsity inducing norm, whose directional derivative exists at every point $\bar{{\bf w}} \in \mathcal{P}$. By combining the result of \cite{konar2017fast} and invoking \cite[Propositon 1]{razaviyayn2013unified}, a link between the first-order properties of  $f_1(\bar{{\bf w}}) + f_2(\bar{{\bf w}})$ and $u(\bar{{\bf w}},\bar{{\bf w}}^{(t)})+ f_2(\bar{{\bf w}})$ can be established about the current SCA iterate $\bar{{\bf w}} = \bar{{\bf w}}^{(t)}$. Together with properties (A1)-(A4), this suffices to obtain the desired convergence claim, by virtue of \cite[Theorem 1]{razaviyayn2013unified}. 

Regarding complexity, it is clear that the main computational cost of Algorithm 1 stems from solving a convex optimization problem of the form (\ref{SCA2}) at every iteration. Since (\ref{SCA2}) does not admit a closed form solution, one should rely on iterative algorithms for solving each SCA subproblem. Instead of resorting to interior-point based methods whose complexity scales unfavorably with the problem dimension, in what follows, we will present two different formulations for~\eqref{SCA2} such that it can be efficiently solved using specialized FOMs. First, we will write the problem in its consensus form which renders it amenable to be solved using the C-ADMM algorithm \cite{boyd2011distributed}. Second, we recast~\eqref{SCA2} in a form that can be efficiently tackled using Nemirovski's Saddle-Point Mirror-Prox method proposed in~\cite{Nemirovski}.  

\section{Proposed Algorithms}\label{prop_alg}
\subsection{Alternating Direction Method of Multipliers}

In this section, we explain how a low-complexity algorithm based on ADMM can be used to solve each subproblem. Define the indicator function of the constraint set $ \mathcal{P} $ as 
\begin{equation}\label{ADMM1}
   \mathbf{\mathit{I}}_{\mathcal{P}}(\bar{{\bf w}}) :=  
    \begin{cases}
                        0,           & \bar{{\bf w}} \in \mathcal{P} \\
                        \infty,      & \text{Otherwise}
    \end{cases}
\end{equation}
Then, each SCA subproblem can be expressed in unconstrained form as
\begin{subequations} \label{ADMM2}
\begin{align}
	 & \underset{\bar{{\bf w}}\in \mathbb{R}^{2N}}{\min} ~   u(\bar{{\bf w}},\bar{{\bf w}}^{(t)}) +  f_2(\bar{{\bf w}}) + \mathbf{\mathit{I}}_{\mathcal{P}}(\bar{{\bf w}}) \\
	 = & \underset{\bar{{\bf w}}\in \mathbb{R}^{2N}}{\min} ~ \sum_{i=1}^{3} g_{i}(\bar{{\bf w}})
	 \end{align}
\end{subequations}
where we have defined $ g_2({\bar{{\bf w}}}) := f_2(\bar{{\bf w}}) $ and $ g_3({\bar{{\bf w}}}) := \mathbf{\mathit{I}}_{\mathcal{P}}(\bar{{\bf w}}) $. Moreover, for notational convenience, we have omitted the dependence of $  u({.,.}) $ on $ \bar{{\bf w}}^{(t)} $ and replaced it with $ g_1({\bar{{\bf w}}}) $. Note that problem (\ref{ADMM2}b) corresponds to the minimization of a sum of convex functions of a common variable $ \bar{{\bf w}} \in \mathbb{R}^{2N}$.

In order to exploit this problem form, we propose to use the consensus form of the ADMM algorithm \cite{boyd2011distributed}, which reduces the problem to iteratively solving a sequence of simpler, specially structured subproblems.  Specifically, consensus ADMM (C-ADMM) creates local copies of the global variable $ \bar{{\bf w}} $ and aims to separately minimize each of the ``local'' cost functions $ g_i(.)$, followed by averaging to enforce consensus. The approach requires evaluating the proximal operator of each of the cost functions at every iteration, which as we will show, can be efficiently performed in parallel. First, we cast (\ref{ADMM2}b) in consensus form as follows
\begin{subequations}\label{ADMM3}
	\begin{align}
	& \underset{\{\bar{{\bf w}}_j\}_{j=1}^{3}}{\min} ~   \sum_{i=1}^{3} g_{i}(\bar{{\bf w}}_i) \\
	& \text{s.t.} \quad \bar{{\bf w}}_1 = \bar{{\bf w}}_2 = \bar{{\bf w}}_3
	\end{align}
\end{subequations}
with the variables $ \bar{{\bf w}}_i  \in \mathbb{R}^{2N}$, $ \forall \;i \in \{1,2,3\} $. Note that each variable corresponds to a copy of the global variable $ \bar{{\bf w}} $, with a consistency constraint for enforcing consensus. We can transform (\ref{ADMM3}) into an unconstrained minimization problem by representing the consensus constraint as the indicator function of the set $ \mathcal{C} := \{(\bar{{\bf w}}_1,\bar{{\bf w}}_2,\bar{{\bf w}}_3)~|~\bar{{\bf w}}_1=\bar{{\bf w}}_2=\bar{{\bf w}} \}$. This allows us to equivalently express (\ref{ADMM3}) as
\begin{equation}\label{ADMM4}
            \underset{\bar{{\bf w}}\in \mathbb{R}^{2N}}{\min} ~ \sum_{i=1}^{3} g_{i}(\bar{{\bf w}}_i) + \mathbf{\mathit{I}}_{\mathcal{C}}(\bar{{\bf w}}_1,\bar{{\bf w}}_2,\bar{{\bf w}}_3)
\end{equation}
where $ \mathbf{\mathit{I}}_{\mathcal{C}}(\bar{{\bf w}}_1,\bar{{\bf w}}_2,\bar{{\bf w}}_3) $ denotes the indicator function of the set $ \mathcal{C} $, and is given by 
\begin{equation}\label{ADMM5}
\mathbf{\mathit{I}}_{\mathcal{C}}(\bar{{\bf w}}_1,\bar{{\bf w}}_2,\bar{{\bf w}}_3) :=  
\begin{cases}
0,           & (\bar{{\bf w}}_1,\bar{{\bf w}}_2,\bar{{\bf w}}_3) \in \mathcal{C} \\
\infty,      & \text{Otherwise}
\end{cases}
\end{equation}
Problem (\ref{ADMM4}) is now in a  form that can be solved using the C-ADMM algorithm. The ADMM updates for (\ref{ADMM4}) are given by 
\begin{subequations}\label{ADMM6}
	\begin{align}
	& \bar{{\bf w}}^{(k+1)}_i \hspace{-1.5cm}&& :=~{\bf prox}_{\rho g_i}(\bar{{\bf w}}_{\text{av}}^{k}-\bar{{\bf v}}^k_i)    \\
	& \bar{{\bf v}}^{(k+1)}_i \hspace{-1.5cm}&& :=~ \bar{{\bf v}}^{(k)}_i + \bar{{\bf w}}^{(k+1)}_i + \bar{{\bf w}}_{\text{av}}^{k+1}
	\end{align}
\end{subequations}
where the superscript $ k = 0,1,\cdots $ is the ADMM iteration counter, the subscript $ i = \{1,2,3\} $ is the subsystem index, $ \bar{{\bf v}}_i $ represents the dual variable for the $ i $-th subsystem and $ \bar{{\bf w}}_{\text{av}} $ denotes the mean of the primal variables of the three subsystems. It should be noted that the primal variables are averaged before computing the dual updates for the three subsystems. Moreover, in (\ref{ADMM6}a), we have defined the proximal operator~\cite{parikh2014proximal} of a convex, proper and closed function $ g: \mathbb{R}^n \rightarrow \mathbb{R}$ as
\begin{equation}\label{ADMM7}
        {\bf prox}_{\rho g}({\bf x}) = \arg \underset{{\bf y}}{\min}~ g({\bf y}) + \frac{1}{2 \rho} \|{\bf y}-{\bf x}\|_2^2
\end{equation} 

\noindent The proximal operators of both $ g_2(\bar{{\bf w}_2}) $ and $ g_3(\bar{{\bf w}_3}) $ exhibit a closed form solution, and hence, they can be computed efficiently. In particular, evaluating the proximal operator of $ g_2(\bar{{\bf w}}) = \lambda\| \bar{{\bf w}} \|_{1,2}$ is equivalent to solving the following problem
\begin{equation}\label{ADMM8}
\begin{aligned}
{\bf prox}_{\rho g_2}({\bf x}) = \arg \underset{\bar{{\bf w}}}{\min}~ & \biggl\{\lambda  \sum\limits_{j=1}^{N}    \|[\bar{{\bf w}}{(j)},\bar{{\bf w}}{(j+N)}]\|_2  \\
&+ \frac{1}{2 \rho} \|\bar{{\bf w}}-{\bf x}\|_2^2 \biggr\}
\end{aligned}
\end{equation}
which can be decomposed to $ N $ separate subproblems, where the solution for each subproblem is given by 
\begin{equation}\label{ADMM9}
({\bf prox}_{\rho g_2}({\bf x}))_j = \bigg( 1 - \frac{\lambda \rho}{\|{\bf x}_j\|_2}\bigg)_{+}{\bf x}_j
\end{equation}
where ${\bf x}_j = [{\bf x}{(j)},{\bf x}{(j+N)}]^T, \forall j \in \{1,\cdots,N\} $.
Furthermore, the proximal operator of $  g_{3}(\bar{\bf w}_3) = \mathbf{\mathit{I}}_{\mathcal{P}}(\bar{{\bf w}}) $ boils down to the Euclidean projection operator for the simple set $ \mathcal{P} $, which can be computed in closed form by separately projecting  each sub-vector $ [\bar{{\bf w}}{(j)},\bar{{\bf w}}{(j+N)}]^T/\sqrt{P_j} $ on the unit $ \ell_2-$norm ball in $\mathbb{R}^2$, $ \forall j \in \{1,\cdots,N\} $.   

Finally, computing the proximal operator of $  g_1(\bar{{\bf w}}) $ requires solving the following unconstrained minimization problem 
\begin{equation}\label{ADMM10}
 \arg\underset{\bar{{\bf w}}}{\min}~ f(\bar{{\bf w}}):=\underset{m}{\max}~\bar{{\bf q}}^{(t)T}_m\bar{{\bf w}}+  \bar{b}_m^{(t)} + \frac{1}{2 \rho} \|\bar{{\bf w}}-{\bf x}\|_2^2
 \end{equation}
Unlike the proximal operators of $ g_2(\bar{{\bf w}}) $ and $ g_3(\bar{{\bf w}}) $, the problem does not possess a closed form solution. Since (\ref{ADMM10}) is quadratic in $ \bar{{\bf w}} $, it can be reformulated in standard quadratic programming (QP) form using the epigraph-based reformulation. However, for massive MIMO scenarios, it will be computationally expensive to solve a QP problem at each ADMM iteration. 

Since (\ref{ADMM10}) is a non-smooth, convex minimization problem, we will replace the piece-wise linear term by the log-sum-exp function,  
\begin{equation}\label{ADMM11}
g_{1}^{\mu}(\bar{{\bf w}}) = \mu \log \bigg(\sum_{m = 1}^{M} \exp\bigg(\frac{\bar{{\bf q}}^{(t)T}_m\bar{{\bf w}} + \bar{b}_m^{(t)}}{\mu}\bigg)\bigg)  -\mu \log M
\end{equation}
which can be interpreted as a differentiable approximation of the piece-wise linear function~\cite{boyd_vandenberghe_2004}. 
Here, $ \mu \in \mathbb{R}^{+}$ is a smoothing parameter that controls the level of smoothness.
We will now show  that the proximal operator of $ g_{1}^{\mu}(\bar{{\bf w}}) $ can be obtained by solving the following \emph{smooth} unconstrained minimization problem
\begin{equation}\label{ADMM12}
\arg\underset{\bar{{\bf w}}}{\min}~f_{\mu}(\bar{{\bf w}}) := g_{1}^{\mu}(\bar{{\bf w}}) + \frac{1}{2 \rho} \|\bar{{\bf w}}-{\bf x}\|_2^2 
\end{equation}
In~\cite{Nesterov2005}, Nesterov established that function $ f_{\mu}(\bar{{\bf w}})  $ has the following properties.
\begin{enumerate}
          \item[(B1)] $ f_{\mu}(\bar{{\bf w}})$ is well defined, strongly convex, and differentiable at all $ \bar{{\bf w}} $ 
          \item[(B2)] $ \nabla f_{\mu}(\bar{{\bf w}})$ is Lipschitz continuous with constant $ L \propto \frac{1}{\mu} $
          \item[(B3)] $ f(\bar{{\bf w}}) - \mu \log M\leq f_{\mu}(\bar{{\bf w}}) \leq  f(\bar{{\bf w}})$
\end{enumerate}
From the third property, it can be easily shown that 
\begin{equation}\label{ADMM_Nest}
      f(\bar{\bf w}) - f^{\star} - \mu \log M \leq f_{\mu}(\bar{\bf w}) - f_{\mu}^{\star}
\end{equation} 
where $ f^{\star} $ and $ f^{\star}_{\mu} $ are the optimal values of~\eqref{ADMM10} and \eqref{ADMM12}, respectively. Let $ \epsilon:=  f(\bar{\bf w}) - f^{\star} $ and $ \epsilon_{\mu} = f_{\mu}(\bar{\bf w}) - f_{\mu}^{\star} $, then it follows from~\eqref{ADMM_Nest} that an $ \epsilon $-optimal solution of~\eqref{ADMM10} corresponds to solving~\eqref{ADMM12} to a numerical accuracy of $ \epsilon_{\mu} = \epsilon - \mu \log M $, i.e., the smooth approximation~\eqref{ADMM12} is solved to a higher degree of accuracy than~\eqref{ADMM10}. Since~\eqref{ADMM12} does not exhibit a closed form solution, one should resort to iterative algorithms. Fortunately, (\ref{ADMM12}) can be solved to an accuracy of $ \epsilon_{\mu} $ in $ O (\sqrt{\frac{L}{\mu}}\log \frac{1}{\epsilon_\mu}) $-iterations using accelerated FOMs~\cite{nesterov1983method,Nesterov2005}. It is clear that the smoothing parameter $ \mu $ effects a trade-off between the tightness of approximation of the non-smooth piece-wise linear function and the convergence rate. In particular, smaller values of $ \mu $ result in less smoothing, and hence, better approximation of the original function. However, larger $ \mu $ provides a more smooth approximation, but results in requiring more iterations to obtain an $ \epsilon_{\mu} $-optimal solution of~(\ref{ADMM12}). We used the following accelerated gradient method for solving~(\ref{ADMM12})  
\begin{algorithm} \label{Alg2}
	\caption{Accelerated gradient scheme}
	\textbf{initialization}: Choose a random starting point $ \bar{{\bf w}}^{(0)} \in \mathbb{R}^{2N}$ and ${\bf y}^{(0)} = \bar{{\bf w}}^{(0)}$  
	\begin{enumerate}
		\item \textbf{for} $ k = 0,1,\cdots $ \textbf{do}
		\item ~~ $ \bar{{\bf w}}^{k+1} = {\bf y}^k - \frac{1}{L}\nabla f({\bf y}^k) $
		\item ~~ $ {\bf y}^{k+1} = \bar{{\bf w}}^{k+1} + \beta(\bar{{\bf w}}^{k+1} - \bar{{\bf w}}^{k}) $
		\item \textbf{end for}
	\end{enumerate}
	\textbf{Until} termination criterion is met
\end{algorithm} 

where $ \beta = \frac{1- \sqrt{\frac{\tau}{L}}}{1+\sqrt{\frac{\tau}{L}}} $, $ \tau $ is the strong convexity parameter of $ f_{\mu}(\bar{{\bf w}})$ and the Lipschitz constant is given in a closed form as follows
\begin{equation}\label{Lip_ADMMM}
	         L = \frac{1}{\rho} + \frac{\|{\bf Q} \|^2_2}{\mu}
\end{equation}
where ${\bf Q} = [\bar{\bf q}_1,\cdots,\bar{\bf q}_M]^T \in \mathbb{R}^{M\times 2N} $.
Note that one can further speedup the convergence rate of the accelerated gradient scheme shown in Algorithm 2 by setting $ \bar{{\bf w}}^{0}= \bar{\bf w}_{\text{av}} $ instead of random initialization. 
 
In~\cite{he20121}, it was established that ADMM converges at a rate of $ O(\frac{1}{k}) $  in an ergodic sense, where $ k $ is the iteration index. Since the proximal operators of $ g_2(.) $ and $ g_3(.) $ are in closed form, the per-iteration complexity is determined by computing the proximal operator of $ g_1(.) $. This requires solving (\ref{ADMM12}) using Algorithm 2 at each ADMM iteration. Meanwhile, the per iteration cost for Algorithm 2 is determined by computing the gradient which requires a matrix-vector multiplication yielding an overall cost of $ O(MN) $ for each iteration. Therefore, it follows that the consensus ADMM can be used to solve each SCA subproblem efficiently. The overall algorithm is given by
\begin{algorithm} \label{Alg3}
	\caption{Consensus ADMM SCA}
\textbf{initialization}: Choose a random starting point $ \bar{{\bf w}}^{(0)} \in \mathcal{P}$ and set $ t := 0 $ \\ 
\textbf{Repeat:}
\begin{itemize}
	\item Compute $ \bar{{\bf w}}^{(t+1)} = \underset{\bar{{\bf w}} \in \mathcal{P}}{\min}~u(\bar{{\bf w}},\bar{{\bf w}}^{(t)}) + f_2(\bar{{\bf w}})$ according to the consensus ADMM updates (\ref{ADMM6}), using Algorithm 2 to solve (\ref{ADMM12})  
	\item set $t = t+1$
\end{itemize}
\textbf{Until} termination criterion is met
\end{algorithm} 

\subsection{Saddle Point Mirror Prox}
We now propose an alternative low complexity method to solve each SCA subproblem. In~\cite{Nemirovski}, Nemirovski  
developed a simple prox-type method for solving problems of the form
\begin{equation}\label{SPMP1}
  \underset{{\bf x} \in \mathcal{X}}\min~\underset{{\bf y} \in \mathcal{Y}}\max ~ \phi(\bf x,y) 
\end{equation}
contingent on $ \phi: \mathbb{R}^n \times \mathbb{R}^m \rightarrow \mathbb{R} $ being a continuous function which is convex in $ {\bf x} \in \mathbb{R}^n$ and concave in $ {\bf y} \in  \mathbb{R}^m $, and the sets $\mathcal{X} \subset \mathbb{R}^n$ and $\mathcal{Y} \subset \mathbb{R}^m$ being ``simple'' convex, compact sets. By Sion's Minimax equality theorem~\cite{sion1958general}, we have that
\begin{equation}\label{SPMP2}
\underset{{{\bf x}} \in \mathcal{X}  }{\min} ~ \underset{{\bf y}\in \mathcal{Y}}{\max} ~ \phi({\bf x},{\bf y})  = \underset{{{\bf x}} \in \mathcal{X}  }{\max} ~ \underset{{\bf y}\in \mathcal{Y}}{\min} ~ \phi({\bf x},{\bf y}), 
\end{equation}
which implies that the optimal solution pair $({\bf x}^*,{\bf y}^*) \in \mathcal{X} \times \mathcal{Y}$ of~\eqref{SPMP1} is a saddle-point of $\phi({\bf x},{\bf y})$, i.e., $({\bf x}^*,{\bf y}^*)$ satisfies
\begin{equation}\label{SPMP3}
\phi({\bf x}^{*},{\bf y})  \leq  \phi({\bf x}^{*},{\bf y}^{*}) \leq \phi({\bf x},{\bf y}^{*}),\;\forall \; ({\bf x,y}) \in  \mathcal{X} \times \mathcal{Y} 
\end{equation} 

We now recast each SCA subproblem~\eqref{SCA2} in the form of \eqref{SPMP1}. 
Since the dual norm of the $ \ell_{1,2} $-norm is the $ \ell_{\infty,2} $-norm~\cite{francis}, it follows that ~\eqref{SCA2} can be written as 
\begin{subequations}\label{SPMP4}
	\begin{align}
	& \underset{\bar{{\bf w}}\in \mathbb{R}^{2N}}{\min} ~ \underset{m \in \mathcal{M}}{\max}~\bar{{\bf q}}^{(t)T}_m\bar{{\bf w}}+  \bar{b}_m^{(t)} + 
	\lambda \underset{\| \bar{\bf s}\|_{\infty,2} \leq 1}{\max} \bar{\bf s}^T\bar{{\bf w}}\\
	& \text{s.t.} \quad \bar{{\bf w}} \in \mathcal{P} 
	\end{align}
\end{subequations}
We now define the vector $ {\bf b}^{(t)} = [\bar{b}_1^{(t)}, \cdots, \bar{b}_M^{(t)}]^T  \in \mathbb{R}^{M}$. By exploiting the fact that maximizing a piece-wise affine function is equivalent to the maximization over the $ M $-dimensional probability simplex, with the maximum attained at one of the simplex's vertices (i.e., a canonical basis vector of $\mathbb{R}^{M}$), \eqref{SPMP4} can be equivalently written as
\begin{subequations}\label{SPMP5}
	\begin{align}
	& \underset{\bar{{\bf w}}\in \mathbb{R}^{2N}}{\min} ~ \underset{\bar{\bf y}\in \bigtriangleup_M}{\max} ~ \bar{\bf y}^T({ {\bf Q}^{(t)}}\bar{{\bf w}}+{ {\bf b}^{(t)}}) + 
	\lambda \underset{\| \bar{\bf s}\|_{\infty,2} \leq 1}{\max} \bar{\bf s}^T\bar{{\bf w}}\\
	& \text{s.t.}  \quad \bar{{\bf w}} \in \mathcal{P}  
	\end{align}
\end{subequations}   
where 
\begin{equation}
\bigtriangleup_M := \big\{\bar{\bf y} \in \mathbb{R}_{+}^{M} ~|~ \sum\limits_{i=1}^{M} \bar{\bf y}(i) = 1\big\}
\end{equation}
denotes the $M$-dimensional probability simplex. Also, we have defined the set 
\begin{equation}
\mathcal{S} := \big\{\bar{\bf s} \in \mathbb{R}^{2N} ~|~ \underset{i}\max ~ \|[ \bar{\bf s}(i), \bar{\bf s}(i{+}N)]^T \|_2 \leq 1, \forall\; i\in [N]\big\} 
\end{equation}
Note that the cost function of~\eqref{SPMP5} is bi-linear, as it is linear (and hence, convex) in $ \bar{\bf w} $ for fixed $ (\bar{\bf y} ,\bar{\bf s}) $, and linear in $ (\bar{\bf y},\bar{\bf s})$ (and hence, concave) for fixed $\bar{\bf w} $. Therefore, as a final step, let us express ($ \ref{SPMP5} $) in a more compact form by defining the matrix $  \bar{\bf Q} = [{\bf Q}^T,\lambda{\bf I}_{2N}]^T$, the vector $ {\bar{\bf x}} = [\bar{\bf y}^T, \bar{\bf s}^T]^T $ and the vector $  \bar{\bf b} = [{\bf b}^T, {\bf 0}_{2N}^T]^T$. Hence, it follows that~\eqref{SPMP5} can be equivalently reformulated as
\begin{equation}\label{SPMP6}
\underset{\bar{{\bf w}} \in \mathcal{P}  }{\min} ~ \underset{\bar{\bf x}\in \bigtriangleup_M \times \mathcal{S}}{\max} ~ \bar{\bf x}^T(\bar{\bf Q}^{(t)}\bar{{\bf w}}+\bar{\bf b}^{(t)}) 
\end{equation}
Defining $ \phi^{(t)}(\bar{\bf w},\bar{\bf x}) := \bar{\bf x}^T(\bar{\bf Q}^{(t)}\bar{{\bf w}}+\bar{\bf b}^{(t)}) $ and the set $ \mathcal{X} :=  \bigtriangleup_M \times \mathcal{S} $,~\eqref{SPMP6} can be finally expressed as
\begin{equation}\label{SPMP7}
\underset{\bar{{\bf w}} \in \mathcal{P}  }{\min} ~ \underset{\bar{\bf x}\in \mathcal{X}}{\max} ~ \phi^{(t)}(\bar{\bf w},\bar{\bf x}) 
\end{equation}
Since $ \phi^{(t)}(\bar{\bf w},\bar{\bf x}) $ is convex in $ \bar{\bf w} \in \mathbb{R}^{2N}$ and concave in $ \bar{\bf x} \in  \mathbb{R}^{M+2N} $ and $ \mathcal{X}, \mathcal{P}  $ are both convex compact sets,~\eqref{SPMP7} exhibits the same form as \eqref{SPMP1}. In \cite{Nemirovski}, Nemirovski proposed the \textit{Saddle Point Mirror Prox} (SP-MP) algorithm for efficiently solving such problems. The method can be considered as a variant of the mirror descent algorithm~\cite{beck2003mirror,bubeck2015convex}, and is outlined below.

First, let $ \Phi_{\mathcal{P}} (\bar{\bf w})$, $ \Phi_{\mathcal{X}} (\bar{\bf x})$  denote ``mirror maps'' for the sets $\mathcal{P}$ and $\mathcal{X}$ (i.e., strongly convex functions capable of exploiting the geometry of the sets), respectively. We define the mirror maps for the sets $ \mathcal{P}, \mathcal{\bigtriangleup}_M ~ \text{and} ~ \mathcal{S}  $ to be $ \Phi_{\mathcal{P}}(\bar{\bf w}) = \| \bar{{\bf w}} \|_2^2  $, $  \Phi(\bar{\bf y}) = \sum\limits_{m=1}^{M} \bar{y}_m \log \bar{y}_m $ and $ \Phi(\bar{\bf s}) = \| \bar{\bf s} \|_2^2 $, respectively.
Upon defining the set $ \mathcal{Z} := \mathcal{P} \times \mathcal{X}$, we construct the following mirror map for $\mathcal{Z}$ 
\begin{equation}
\begin{aligned}
\Phi_{\mathcal{Z}}(\bar{\bf z}) &= \Phi_{\mathcal{Z}}(\bar{\bf w}, \bar{\bf x}) \\
&= \Phi_{\mathcal{W}} (\bar{\bf w}) + \Phi_{\mathcal{X}} (\bar{\bf x})\\
& = \| \bar{{\bf w}} \|_2^2 + \sum\limits_{m=1}^{M} \bar{ y}_m \log\bar{y}_m + \| \bar{\bf s} \|_2^2 
\end{aligned}
\end{equation}
The mirror map serves as a distance generating function, which allows us to define the Bregman divergence associated with $\Phi_{\mathcal{Z}}(\bar{\bf z})$ as
\begin{equation}\label{BD}
D_{\Phi}(\bar{\bf z},\bar{\bf z}'):= \Phi(\bar{\bf z}')-\Phi(\bar{\bf z})- \nabla\Phi(\bar{\bf z})^T(\bar{\bf z}-\bar{\bf z}'), \forall \, \bar{\bf z},\bar{\bf z}' \in \mathcal{Z}
\end{equation}
 Furthermore, defining $ \beta := \max\{\beta_{ij}\} $, for $ i,j \in \{1,2\} $, and the step size $ \alpha = \frac{1}{2\beta} $, the SP-MP algorithm is given by the steps shown in Algorithm 1.
\begin{algorithm} \label{A1}
	\caption{Saddle Point Mirror-Prox}
	\textbf{initialization}: Define $ \bar{\bf z}_t = [\bar{{\bf w}}_t^T,\bar{{\bf x}}_t^T] $, 
	$ \bar{\bf r} = [\bar{{\bf u}}_t^T,\bar{{\bf v}}_t^T] $, 
	$ \Psi({\bar{\bf z}}_t)  = [\nabla_{\bar{{\bf w}}} \phi(\bar{{\bf w}}_t,\bar{{\bf x}}_t)^T,-\nabla_{\bar{{\bf x}}} \phi(\bar{{\bf w}}_t,\bar{{\bf x}}_t)^T$], and
	$\Psi(\bar{{\bf r}})  = [\nabla_{\bar{{\bf u}}} \phi(\bar{{\bf u}}_t,\bar{{\bf v}}_t)^T,-\nabla_{\bar{\bf v}} \phi(\bar{{\bf u}}_t,\bar{{\bf v}}_t)^T$] for $ t \geq 0 $, starting with feasible $ \bar{\bf z}_0 $   \\ 
	\textbf{Repeat:}
	\begin{enumerate}
		\item $\nabla \Phi(\bar{\bf r}'_{t+1}) = \nabla \Phi(\bar{\bf z}_t) - \alpha \Psi(\bar{\bf z}_{t}) $
		\item $ \bar{\bf r}'_{t+1} = \nabla \Phi^{-1}(\nabla \Phi(\bar{\bf z}_t) - \alpha \Psi(\bar{\bf z}_{t}))$
		\item $  \bar{\bf r}_{t+1}  = \arg \min_{\bar{\bf z} \in \mathcal{Z}} \mathcal{D}_{\Phi}({\bar{\bf z}},{\bf r}'_{t+1})$
		\item $\nabla \Phi(\bar{\bf z}'_{t+1}) = \nabla \phi({\bar{\bf z}}_t) - \alpha \Psi({\bf r}_{t+1}) $
		\item $ \bar{\bf z}^{'}_{t+1} = \nabla \Phi^{-1}(\nabla \Phi({\bar{\bf z}}_t) - \alpha \Psi({{\bf r}}_{t+1}))$
		\item $ {\bf {z}}_{t+1}  = \arg \min_{{\bf {z}} \in \mathcal{Z}} D_{\Phi}({\bar{\bf z}},{\bar{\bf z}'}_{t+1})$
		\item set $t = t+1$
	\end{enumerate}
	\textbf{Until} termination criterion is met
\end{algorithm} 

Note that the quantities $ \nabla \Phi(\bar{\bf z}) $  and $ \nabla^{-1} \Phi(\bar{\bf z}) $ can easily be computed as follows
\begin{eqnarray} \label{SPMP9}
\nabla \Phi(\bar{\bf z}) &{=}& \hspace{-0.25cm}[\bar{\bf w},1+\log{y}_1,\cdots,1+\log{y}_M,\bar{\bf s}] \\
\nabla^{-1} \Phi(\bar{\bf z}) &{=}& \hspace{-0.25cm}[\bar{\bf w},\exp({y}_1 -1),\cdots,\exp({y}_M -1),\bar{\bf s}]
\end{eqnarray}
Moreover, by using the definition of the Bregman Divergence in (\ref{BD}), the non-Euclidean projection problem in Algorithm 4 can be written as
\begin{eqnarray}\label{SPMP10}
\underset{\bar{\bf z} \in \mathcal{Z}} \min D_{\Phi}(\bar{\bf z},\bar{\bf z}') {=} 
\min_{\substack{\bar{{\bf w}} \in \mathcal{W}\\\bar{\bar{\bf y}} \in \bigtriangleup_M \\ {\bar{\bf s}} \in \mathcal{S}}} \frac{1}{2}\| \bar{{\bf w}} - \bar{{\bf w}}' \|_2^2 +  \frac{1}{2}\| {\bar{\bf s}} - {\bar{\bf s}}' \|_2^2  \nonumber \\
{+}\sum\limits_{m=1}^{M} \bar{y}_m \log\frac{\bar{y}_m}{\bar{y}'_m} - \sum\limits_{m=1}^{M}(\bar{y}_m - \bar{y}'_m) 
\end{eqnarray}
The above problem can be resolved into three distinct projection problems as follows
\begin{subequations}\label{SPMP11}
	\begin{align}
	\underset{\bar{{\bf w}} \in \mathcal{W}}\min ~ &\frac{1}{2}\| \bar{{\bf w}} - \bar{{\bf w}}' \|_2^2, \\
	\underset{{\bar{\bf s}} \in \mathcal{S}}\min ~ &\frac{1}{2}\| {\bar{\bf s}} - {\bar{\bf s}}' \|_2^2, \\
	\underset{{\bar{\bf y}} \in \bigtriangleup_M}\min ~ &\sum\limits_{m=1}^{M} { y}_m \log\frac{\bar{y}_m}{\bar{y}'_m} - \sum\limits_{m=1}^{M}(\bar{y}_m - \bar{y}'_m)
	\end{align}
\end{subequations} 
Note that problems (\ref{SPMP11}a) and (\ref{SPMP11}b) are Euclidean projection onto $ \mathcal{P} $ and $ \mathcal{S} $, respectively. It is easy to see that computing the projection for the set $ \mathcal{S}$ reduces  to  separately projecting each sub-vector $ {\bar{\bf s}}_j=  [s(j), s(j+N)]^T $ on the unit $ l_2$-ball, where $ j \in [N] $. On the other hand, the projection on the $ M $-dimensional probability simplex has a simple closed form solution~\cite{bubeck2015convex} given by 
\begin{equation}
{ \bar{\bf y}} = \left\{\begin{array}{lr}
{ \bar{\bf y}}', & {\bar{\bf y}}' \in \bigtriangleup_M \\
\frac{ { \bar{\bf y}}'}{\lVert  {\bar{\bf y}}^{'} \rVert_1}, &  \text{otherwise}
\end{array}\right\}
\end{equation} 
Finally, the step size $ \alpha = \frac{1}{2L} $ can be obtained from~\eqref{SPMP9} by noticing that $ \beta_{11} = \beta_{22} =0 $ and $ \beta_{12} = \beta_{21} = L $, where the Lipschitz constant  $ L $  is given by
\begin{equation}
L = \max(\underset{m}\max(\|\bar{\bf q}^{(t)}_m\|_2),\lambda)
\end{equation}
The steps of the SP-MP algorithm can now be applied to solve~\eqref{SCA2} to obtain the optimal solution 
$ \bar{{\bf w}}^{\star(t)} $ at the $ t $-th iteration.
Then the SCA iterative algorithm is used to get the final solution $ \bar{{\bf w}}^{\star} $ that corresponds to the desired beamforming vector. Summarizing, the overall algorithm is given by the following steps 
\begin{algorithm} \label{A2}
	\caption{SP-MP SCA}
	\textbf{initialization}: Randomly generate a feasible starting point $ \bar{{\bf w}}^{(0)} \in \mathcal{P} $, set $ t:=0 $ \\
	\quad \textbf{Repeat:}
	\begin{itemize}
		\item compute $  \bar{{\bf w}}^{(t+1)} $ using SP-MP (Algorithm 4).
		\item set $ t:=t+1 $
	\end{itemize}
	\textbf{Until} termination criterion is met 
\end{algorithm} 

It is worth mentioning that Nemirovski established a convergence analysis in terms of the iterates generated by the algorithm, where an iteration upper bound of $ O(\frac{1}{\epsilon}) $ is derived to guarantee an $ \epsilon $-optimal solution of~\eqref{SPMP7}. Moreover, as all projections exhibit closed form solutions, the cost of each iteration of the SP-MP algorithm is determined by computing the saddle-point-operator $ \Psi(\bar{\bf z}) $, which requires only $ O(MN) $ operations that result from matrix-vector multiplications. This enables obtaining high quality solutions at low complexity, as we will see. 

\subsection{Choice of $ \lambda $}
In the aforementioned algorithms, the sparsity-inducing parameter $ \lambda $ was assumed to be constant in both the C-ADMM SCA and the SP-MP SCA algorithms described in Algorithms $ 3 $ and $ 5  $, respectively. 
The final solution $ \bar{{\bf w}}^{\star} $ obtained from either algorithm is sparse, where the degree of sparsity depends on the strength of the regularizer $ \lambda $. A bisection search is used to get the $ \lambda $ that yields the desired set of antennas required for transmission. For a given upper bound $ \lambda_{UB} $ and lower bound $ \lambda_{LB} $, we set $ \lambda = \frac{\lambda_{UB} - \lambda_{LB}}{2} + \lambda_{LB}$ and run the desired SCA algorithm (using SP-MP  or C-ADMM). Let $ S $ denote the number of non zero entries in the beamforming vector, i.e., $S = \text{card}({\bf w}^{\star})$. If $ S = K $, then we are essentially done: we just run the SP-MP SCA algorithm one more time for the selected antennas to obtain the final beamforming vector. Otherwise, if $ S > K $, set $\lambda_{LB} = \lambda  $, while if $ S< K $, set $ \lambda_{UB} = \lambda $, and repeat until $ S = K $. The overall algorithm required to solve the joint problem is described below
\begin{algorithm} \label{A6}
	\caption{Bisection search}
	\textbf{initialization}: Set $ \lambda_{UB} = u $, $ \lambda_{LB} = l $ 
		\begin{enumerate}
			\item \textbf{while} $ S  \neq K$ \textbf{do}
		    \item ~~ Compute $ \lambda = \frac{\lambda_{UB}+\lambda_{LB}}{2} $ 
			\item ~~ Compute $ \bar{{\bf w}}^{\star} $ using the SP-MP or C-ADMM SCA 
			\item ~~~~~ \textbf{if} $ S > K $ 
			\item ~~~~~~~~ set $\lambda_{LB} = \lambda$
			\item ~~~~~ \textbf{else}
			\item ~~~~~~~~ set $\lambda_{UB} = \lambda$
			\item ~~~~~ \textbf{endif}
			\item \textbf{end while}
		\end{enumerate}
\end{algorithm}

\section{Simulation Results}\label{SR}
To evaluate the performance of our proposed low-complexity FOMs SCA algorithms, we consider two scenarios for the problem; the traditional single group multicasting scenario where the number of users is greater than the number of antennas, i.e., $ N \leq M $, and the massive MIMO scenario where the number of antennas at the BS is much greater than the number of users, i.e., $ N >> M $. For both settings, the downlink channels are modeled as 
\begin{equation}\label{SR1}
{\bf h}_m^H = \sqrt{\frac{N}{L_m}} \sum\limits_{l=1}^{L_m} \alpha_m^{(l)}{\bf a}^H_{t}(\theta_{t_m}^{(l)}), \forall m \in \mathcal{M}
\end{equation}
where $ L_m \sim \mathcal{U}[5,\cdots,20]$ is the number of scattering paths between the BS and the $ m $-th user, $ \alpha_m^{(l)} \sim \mathcal{CN}(0,1)$ is the complex gain of the $ l $-th path, $ {\bf a}_{t}(.) $ is the transmit array steering vector which depends on the antenna array geometry, and $ \theta_{t_m}^{(l)} \sim \mathcal{U}[-\pi/2,\pi/2]$ denotes the azimuth angle of departure of the $ l $-th path. Assuming the BS is equipped with a uniform linear array, then
\begin{equation} \label{SR2}
{\bf a}_t(\theta) = [1,e^{(i2\pi d/\gamma)\sin(\theta)}, \cdots,e^{(i2\pi (N-1) d/\gamma)\sin(\theta)}]^T
\end{equation}
where $ \gamma $ is the carrier wavelength and $ d = \gamma/2$ is the distance between the adjacent antennas. The noise variance $ \sigma_m^2  $ was set to $ 1 $, $ \forall m \in \mathcal{M} $.             

The C-ADMM SCA and the SP-MP SCA algorithms were implemented in MATLAB. Within the SCA loop, the solution from the previous iteration was used to warm-start the next iteration. For C-ADMM SCA, we set the smoothing parameter for the point-wise linear approximation to $ \mu = 1e^{-2} $, and the ADMM penalty parameter $ \rho = 0.1 $. For the accelerated gradient scheme, the step size was set to $ \frac{1}{L} $ where $ L = \frac{1}{\rho} + \frac{\|\bar{\bf Q} \|_2^2}{\mu} $, and the $ \epsilon_\mu $-accuracy was set to $ 1e^{-5} $. For both algorithms, the $ \epsilon $-accuracy was set to $ 1e^{-5} $ while the maximum number of iterations for solving each sub-problem was set to $1000$. The SCA algorithms were all initialized from the same starting point and run for a maximum of $ 15 $ iterations. In all simulations, we have performed $100$ Monte-Carlo trials on a Windows desktop with $ 6 $ Intel i$ 7 $ cores and $ 16 $GB of RAM.

\subsection{Traditional Multicasting Scenario}
\begin{figure}[!t]
	\centering
	\includegraphics[width=3in]{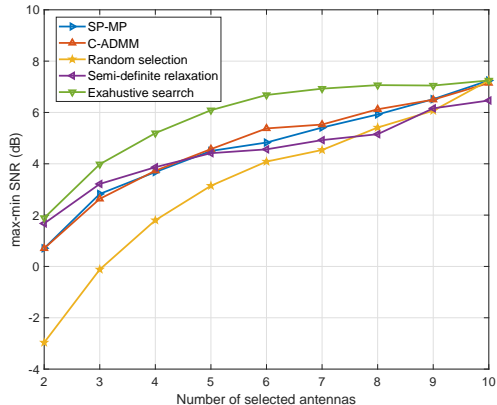}
	\caption{max-min SNR vs number of selected antennas with $ N = 10 $ and $ M=50 $}
	\label{SNR_trad}
\end{figure}

\begin{figure}[!t]
	\centering
	\includegraphics[width=3in]{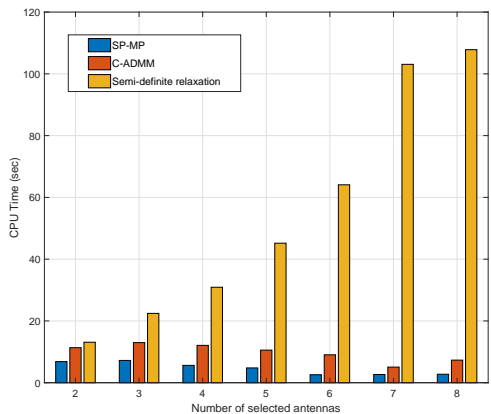}
	\caption{CPU time vs the number of selected antennas with $ N = 10 $ and $ M = 50 $}
	\label{time_trad}
\end{figure}

We first consider a traditional multicast scenario where $ N \leq M $. The simulation setup included a BS with $ N = 10 $ broadcasting a common message to $ M = 50 $ users. In contrast to the massive MIMO scenario where PAPCs are considered, we replaced PAPCs with a sum power constraint and set $ P = 10 $. Moreover, it was empirically found that, setting $ \lambda_{UB} = 1 $ and $ \lambda_{LB} = 0 $ is sufficient to cover the required range of $ \lambda $ for the binary search to get the optimal $ \lambda $ required for the desired sparse solution. In order to assess the performance of our proposed algorithms for this scenario, we used the modeling language YALMIP~\cite{lofberg2004yalmip}, that uses SeDuMi as a solver, to implement the SDP-based algorithm presented in~\cite{mehanna2013joint}. Moreover, to obtain the optimal set of selected antennas and their corresponding beamforming vector, we run exhaustive search over all possible patterns and use it as a performance benchmark. Note that the SP-MP SCA was used to obtain the beamforming vector for each set, where all patterns are initialized from the same starting point. The running time results are depicted in Figure \ref{time_trad}. We observe that the Nemirovski SCA approach, which uses the SP-MP algorithm to solve each subproblem, is up to $ 30 $ times faster for some value of $ K $ relative to the state-of-art SDR. In addition, SP-MP SCA is two times faster than the C-ADMM SCA. On the other hand, in terms of max-min SNR, Figure \ref{SNR_trad} shows that SDR performs slightly better than our SCA algorithms at lower values of $ K $. However, for $ K \geq 5 $, C-ADMM SCA exhibits the highest max-min SNR compared to SDR and SP-MP SCA. Although SP-MP SCA is always the second best either after SDR at lower values of $ K $ or after the C-ADMM SCA at higher values of $ K $, it still remains the best in terms of computational complexity.

\subsection{Massive MIMO Multicasting Scenario}
\begin{figure}[!t]
	\centering
	\includegraphics[width=3in]{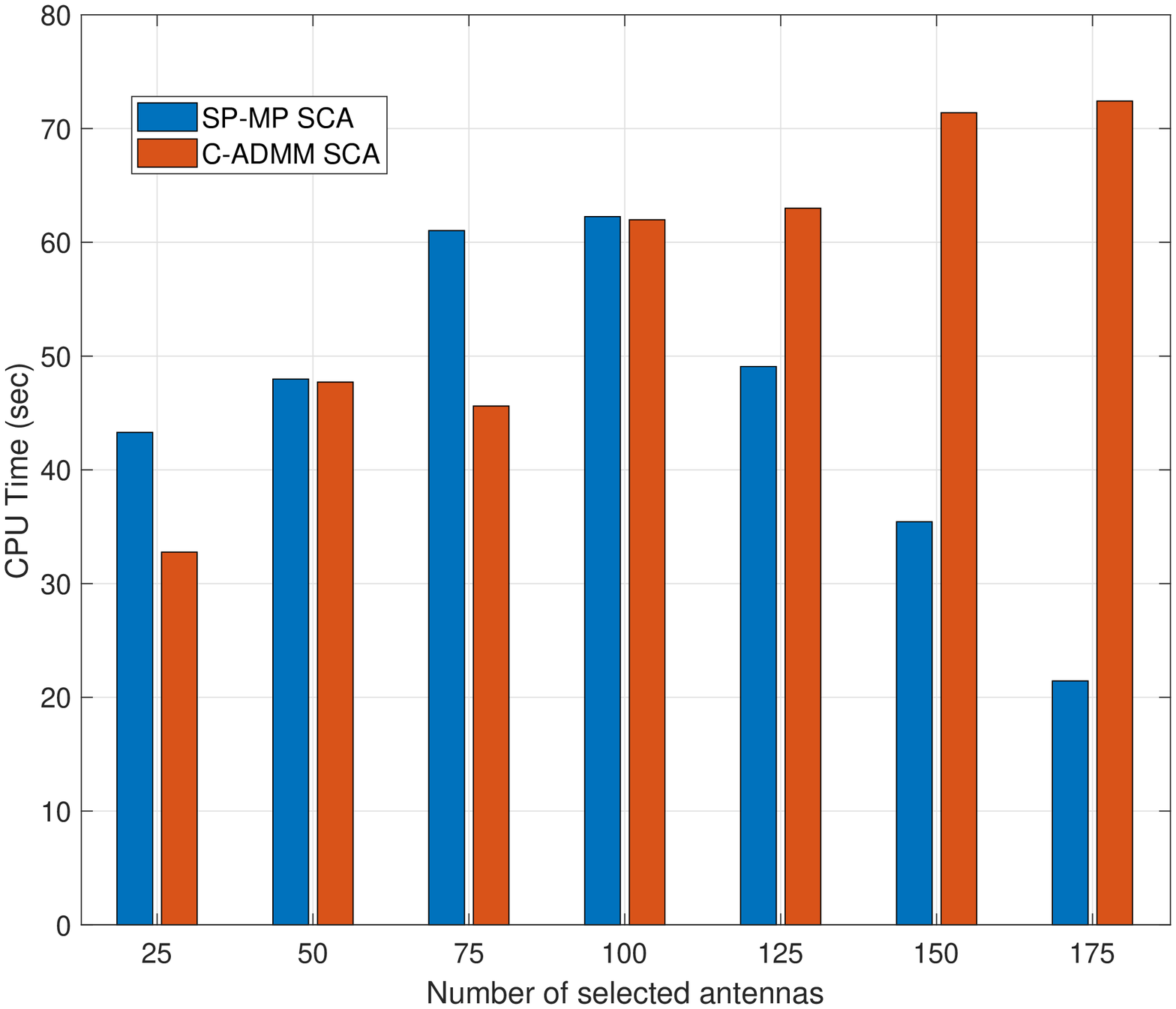}
	\caption{ CPU time vs. the number of selected antennas, with $ N = 200 $ and $ M=50 $}
	\label{time_Massive}
\end{figure}

\begin{figure}[!t]
	\centering
	\includegraphics[width=3in]{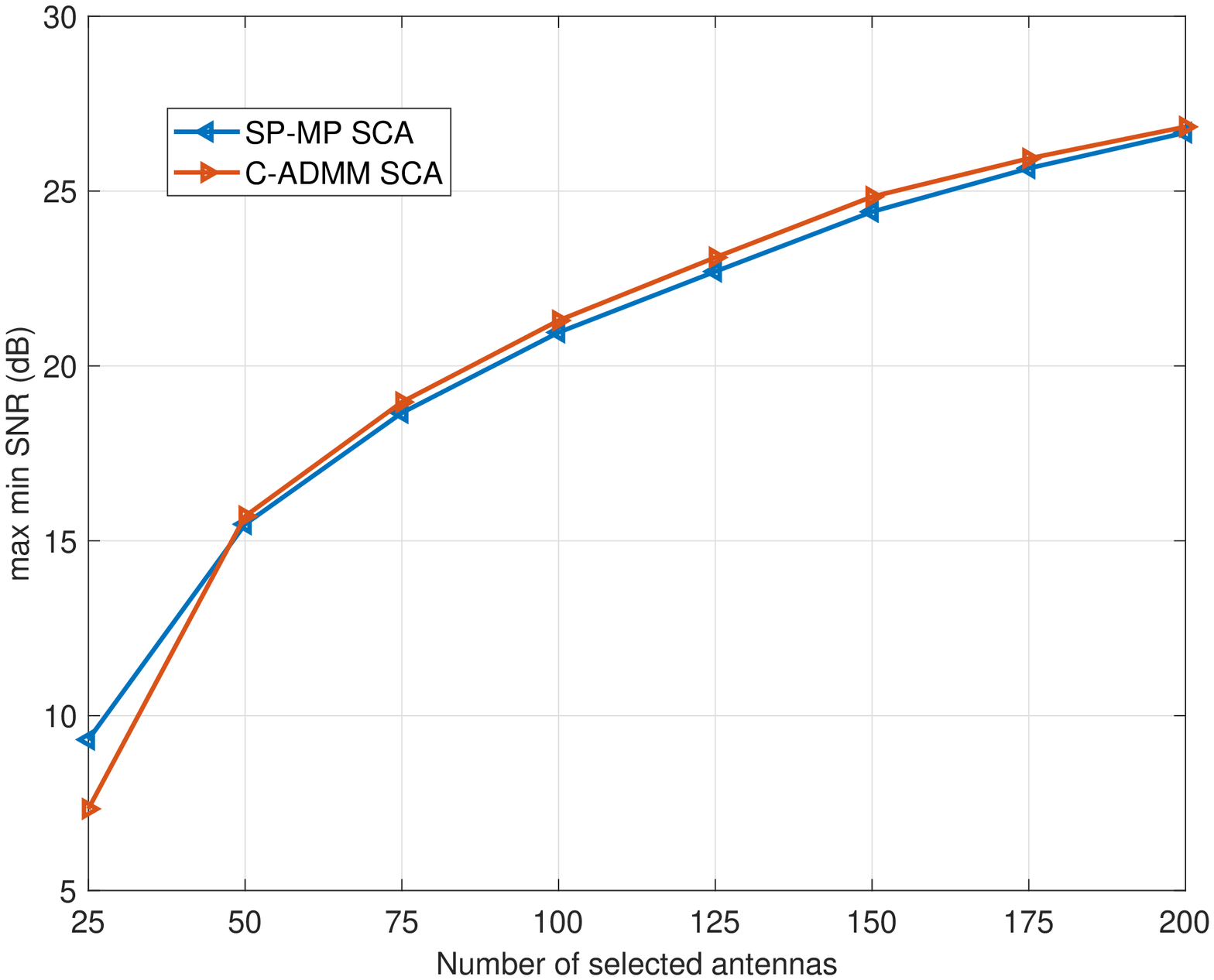}
	\caption{max-min SNR vs number of selected antennas with $ N = 200 $ and $ M=50 $}
	\label{SNR_Massive}
\end{figure}

We carried out simulations for massive MIMO multicasting, where $ N = 200 $ and $ M = 50 $. The number of selected antennas $ K $ is varied from $ 25 $ to $ 200 $. In this setting, we set $ \lambda_{UB} = 2 $ and $ \lambda_{LB} = 0 $. In addition, the power budget for each antenna was set to be $ P_j = 0 .5$, $ \forall j \in \{1,\cdots,N\} $. Figure~\ref{time_Massive} depicts the number of iterations required for each SCA algorithm to reach the final sparse solution which corresponds to the selected set of antennas, and their corresponding beamforming vector. We observe that the C-ADMM SCA obtains the desired sparse solution in much fewer iterations compared to the SP-MP SCA. This renders the C-ADMM SCA considerably cheaper than the SP-MP, for $ K \leq 100 $. However, for $ K > 100 $, although C-ADMM SCA still provides savings in the number of iterations needed to find the desired sparse solution, Figure~\ref{time_Massive} shows that SP-MP SCA performs better in terms of run time. This can be  attributed to the fact that C-ADMM requires applying the accelerated gradient method at each iteration to compute the prox operator of the point-wise linear function, however, all the expressions of the SP-MP are in closed form. Therefore, as the gap in the number of iterations shrinks, the SP-MP can perform faster than the C-ADMM. Regarding solution quality, Figure~\ref{SNR_Massive} indicates that both algorithms can attain approximately the same max min SNR. 
\begin{figure}[!t]
	\centering
	\includegraphics[width=3in]{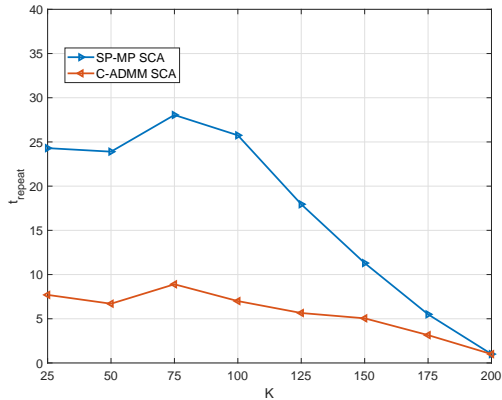}
	\caption{$ t_{\text{repeat}} $ vs $ K $, where $ t_{\text{repeat}} $ represents the number of bisection steps, for $ N = 200 $ and $ M=50 $}
	\label{Itr_Massive}
\end{figure} 

\begin{figure}[!t]
	\centering
	\includegraphics[width=3in]{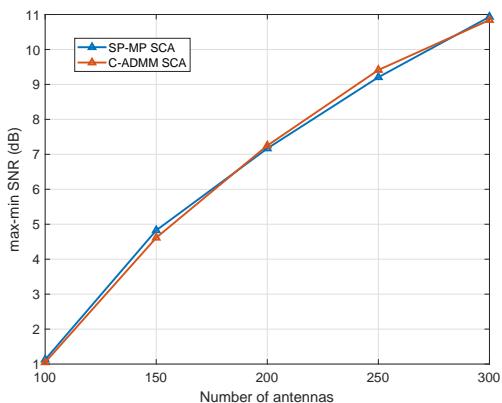}
	\caption{max-min SNR vs number of antennas with $ K = N/10 $ and $ M=50 $}
	\label{SNR_Massive_RF}
\end{figure}
\begin{figure}[!t]
	\centering
	\includegraphics[width=3in]{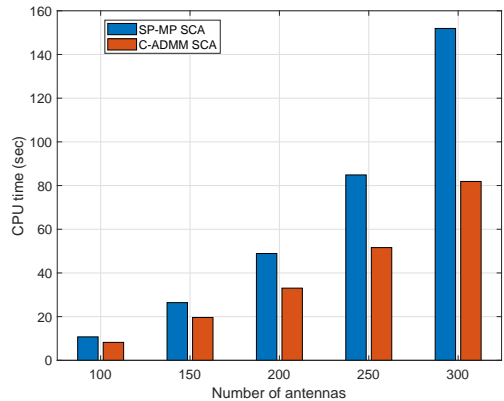}
	\caption{CPU time vs. the number of antennas, with $ K = N/10 $ and $ M=50 $}
	\label{time_Massive_RF}
\end{figure}
\begin{figure}[!t]
	\centering
	\includegraphics[width=3in]{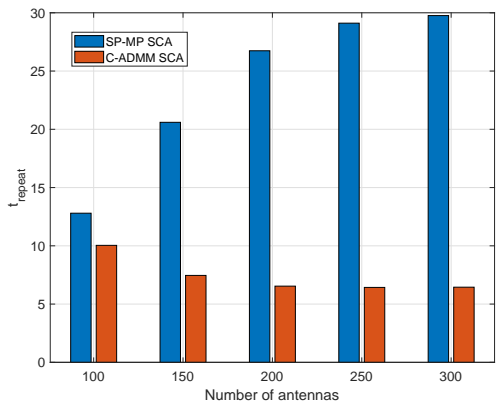}
	\caption{$ t_{\text{repeat}} $ vs $ N $, where $ t_{\text{repeat}} $ represents the number of bisection steps needed to get to the desired sparse solution, for $ K = N/10 $ and $ M=50 $}
	\label{Itr_Massive_RF}
\end{figure}

We also carried out one more comprehensive experiment for the massive MIMO setting. The number of antennas at the BS $ N $ was varied from $ 100 $ to $ 300 $, and the number of available RF chains was set to $ 10\% $ of the number of antennas at the BS, i.e., $ K = 10\%$ of $N$. The number of users was set to $50$. The algorithm parameters were set as explained before. All results for this scenario were obtained by averaging over $ 100 $ channel realizations for each value of $ N $. Figure~\ref{SNR_Massive_RF} shows that the max-min SNR obtained by the two algorithms is approximately the same, on average. In terms of execution time, it is observed that the C-ADMM SCA algorithm is cheaper than SP-MP SCA. Figure~\ref{time_Massive_RF} shows that, as the number of antennas $ N $ increases, C-ADMM becomes faster than SP-MP. This is attributed to the ability of C-ADMM to reach the desired sparse solutions much faster than SP-MP, see Figure~\ref{Itr_Massive_RF}.  This is because C-ADMM is more agressive in zeroing out elements of the beamforming vector, thus requiring fewer bisection steps. 

\section{Conclusion}\label{Conc}
The problem of max-min fair multicasting with antenna selection was studied. The goal is to select a subset of antennas that maximizes the minimum received SNR among all users after appropriate beamforming. A SCA approach was developed for the purpose of obtaining a high-quality, approximate solution for the problem at low-complexity. The approach relies on iteratively approximating the non-convex problem via a  sequence of convex problems. Using group sparsity and its dual norm representation, each non-smooth, convex subproblem was equivalently reformulated into two different forms, on which two specialized FOMs were applied. In particular, one reformulation is well suited to be solved using the C-ADMM, while the other can be efficiently handled using Nemirovski's SP-MP method. Both SCA algorithms have shown a superior performance over the state-of-art SDR based approach. In addition, due to the computational efficiency of the proposed algorithms, they can be easily scaled to large size problems, i.e., massive MIMO settings with hundreds of transmit antennas. Two simulation scenarios were considered in detail: traditional multicasting, where $ N << M $, and massive MIMO with $ N >> M $. Results revealed that the proposed methods provide substantial computational savings and a higher attainable max-min SNR (and thus multicast rate) compared to the SDR one. 

%
\section*{Acknowledgment}

The authors would like to thank Prof. Mingyi Hong, who provided useful feedback on an early version of this work \cite{ibrahim2018prox} as part of a course project.

\bibliographystyle{IEEEtran}
\bibliography{IEEEabrv,refrences}

\begin{thebibliography}{10}
\providecommand{\url}[1]{#1}
\csname url@samestyle\endcsname
\providecommand{\newblock}{\relax}
\providecommand{\bibinfo}[2]{#2}
\providecommand{\BIBentrySTDinterwordspacing}{\spaceskip=0pt\relax}
\providecommand{\BIBentryALTinterwordstretchfactor}{4}
\providecommand{\BIBentryALTinterwordspacing}{\spaceskip=\fontdimen2\font plus
\BIBentryALTinterwordstretchfactor\fontdimen3\font minus
  \fontdimen4\font\relax}
\providecommand{\BIBforeignlanguage}[2]{{%
\expandafter\ifx\csname l@#1\endcsname\relax
\typeout{** WARNING: IEEEtran.bst: No hyphenation pattern has been}%
\typeout{** loaded for the language `#1'. Using the pattern for}%
\typeout{** the default language instead.}%
\else
\language=\csname l@#1\endcsname
\fi
#2}}
\providecommand{\BIBdecl}{\relax}
\BIBdecl

\bibitem{osseiran2014scenarios}
A.~Osseiran, F.~Boccardi, V.~Braun, K.~Kusume, P.~Marsch, M.~Maternia,
  O.~Queseth, M.~Schellmann, H.~Schotten, H.~Taoka \emph{et~al.}, ``Scenarios
  for 5{G} mobile and wireless communications: the vision of the {METIS}
  project,'' \emph{IEEE Communications Magazine}, vol.~52, no.~5, pp. 26--35,
  2014.

\bibitem{mobile2011global}
C.~V. Mobile, ``Global mobile data traffic forecast update 2010-2015,''
  \emph{Cisco White Paper}, 2011.

\bibitem{larsson2014massive}
E.~G. Larsson, O.~Edfors, F.~Tufvesson, and T.~L. Marzetta, ``Massive {MIMO}
  for next generation wireless systems,'' \emph{IEEE Communications Magazine},
  vol.~52, no.~2, pp. 186--195, 2014.

\bibitem{boccardi2014five}
F.~Boccardi, R.~W. Heath, A.~Lozano, T.~L. Marzetta, and P.~Popovski, ``Five
  disruptive technology directions for {5G},'' \emph{IEEE Communications
  Magazine}, vol.~52, no.~2, pp. 74--80, 2014.

\bibitem{ngo2013energy}
H.~Q. Ngo, E.~G. Larsson, and T.~L. Marzetta, ``Energy and spectral efficiency
  of very large multiuser {MIMO} systems,'' \emph{IEEE Transactions on
  Communications}, vol.~61, no.~4, pp. 1436--1449, 2013.

\bibitem{sidiropoulos2006transmit}
N.~D. Sidiropoulos, T.~N. Davidson, and Z.-Q. Luo, ``Transmit beamforming for
  physical-layer multicasting,'' \emph{IEEE Transactions on Signal Processing},
  vol.~54, no.~6, pp. 2239--2251, 2006.

\bibitem{lecompte2012evolved}
D.~Lecompte and F.~Gabin, ``Evolved multimedia broadcast/multicast service
  (e{MBMS}) in {LTE}-advanced: overview and rel-11 enhancements,'' \emph{IEEE
  Communications Magazine}, vol.~50, no.~11, pp. 68--74, 2012.

\bibitem{karipidis2008quality}
E.~Karipidis, N.~D. Sidiropoulos, and Z.-Q. Luo, ``Quality of service and
  max-min fair transmit beamforming to multiple cochannel multicast groups,''
  \emph{IEEE Transactions on Signal Processing}, vol.~56, no.~3, pp.
  1268--1279, 2008.

\bibitem{xiang2014massive}
Z.~Xiang, M.~Tao, and X.~Wang, ``Massive {MIMO} multicasting in noncooperative
  cellular networks,'' \emph{IEEE Journal on Selected Areas in Communications},
  vol.~32, no.~6, pp. 1180--1193, 2014.

\bibitem{zhou2015joint}
H.~Zhou and M.~Tao, ``Joint multicast beamforming and user grouping in massive
  {MIMO} systems,'' in \emph{IEEE International Conference on Communications
  (ICC)}, London, UK, June 2015, pp. 1770--1775.

\bibitem{sadeghi2015multi}
M.~Sadeghi and C.~Yuen, ``Multi-cell multi-group massive {MIMO} multicasting:
  {A}n asymptotic analysis,'' in \emph{IEEE Global Communications Conference
  (GLOBECOM)}, CA, USA, Dec. 2015, pp. 1--6.

\bibitem{yu2007transmitter}
W.~Yu and T.~Lan, ``Transmitter optimization for the multi-antenna downlink
  with per-antenna power constraints,'' \emph{IEEE Transactions on Signal
  Processing}, vol.~55, no.~6, pp. 2646--2660, 2007.

\bibitem{christopoulos2014multicast}
D.~Christopoulos, S.~Chatzinotas, and B.~Ottersten, ``Multicast multigroup
  beamforming under per-antenna power constraints,'' in \emph{IEEE
  International Conference on Communications (ICC)}, Australia, June 2014, pp.
  4704--4710.

\bibitem{christopoulos2014weighted}
------, ``Weighted fair multicast multigroup beamforming under per-antenna
  power constraints,'' \emph{IEEE Transactions on Signal Processing}, vol.~62,
  no.~19, pp. 5132--5142, 2014.

\bibitem{christopoulos2015multicast}
------, ``Multicast multigroup beamforming for per-antenna power constrained
  large-scale arrays,'' in \emph{IEEE 16th International Workshop on Signal
  Processing Advances in Wireless Communications (SPAWC)}, Sweden, June 2015,
  pp. 271--275.

\bibitem{marks1978general}
B.~R. Marks and G.~P. Wright, ``A general inner approximation algorithm for
  nonconvex mathematical programs,'' \emph{Operations Research}, vol.~26,
  no.~4, pp. 681--683, 1978.

\bibitem{razaviyayn2013unified}
M.~Razaviyayn, M.~Hong, and Z.-Q. Luo, ``A unified convergence analysis of
  block successive minimization methods for nonsmooth optimization,''
  \emph{SIAM Journal on Optimization}, vol.~23, no.~2, pp. 1126--1153, 2013.

\bibitem{beck2010sequential}
A.~Beck, A.~Ben-Tal, and L.~Tetruashvili, ``A sequential parametric convex
  approximation method with applications to nonconvex truss topology design
  problems,'' \emph{Journal of Global Optimization}, vol.~47, no.~1, pp.
  29--51, 2010.

\bibitem{7874154}
E.~{Chen} and M.~{Tao}, ``{ADMM}-based fast algorithm for multi-group multicast
  beamforming in large-scale wireless systems,'' \emph{IEEE Transactions on
  Communications}, vol.~65, no.~6, pp. 2685--2698, June 2017.

\bibitem{konar2017fast}
A.~Konar and N.~D. Sidiropoulos, ``Fast approximation algorithms for a class of
  non convex {QCQP} problems using first-order methods,'' \emph{IEEE
  Transactions on Signal Processing}, vol.~65, no.~13, pp. 3494--3509, 2017.

\bibitem{8446006}
R.~{Jiang}, H.~{Liu}, and A.~M. {So}, ``{LPA-SD}: An efficient first-order
  method for single-group multicast beamforming,'' in \emph{IEEE 19th
  International Workshop on Signal Processing Advances in Wireless
  Communications (SPAWC)}, Greece, June 2018, pp. 1--5.

\bibitem{sanayei2004antenna}
S.~Sanayei and A.~Nosratinia, ``Antenna selection in {MIMO} systems,''
  \emph{IEEE Communications Magazine}, vol.~42, no.~10, pp. 68--73, 2004.

\bibitem{nai2010beampattern}
S.~E. Nai, W.~Ser, Z.~L. Yu, and H.~Chen, ``Beampattern synthesis for linear
  and planar arrays with antenna selection by convex optimization,'' \emph{IEEE
  Transactions on Antennas and Propagation}, vol.~58, no.~12, pp. 3923--3930,
  2010.

\bibitem{dua2006receive}
A.~Dua, K.~Medepalli, and A.~J. Paulraj, ``Receive antenna selection in {MIMO}
  systems using convex optimization,'' \emph{IEEE Transactions on Wireless
  Communications}, vol.~5, no.~9, pp. 2353--2357, 2006.

\bibitem{gao2015massive}
X.~Gao, O.~Edfors, F.~Tufvesson, and E.~G. Larsson, ``Massive {MIMO} in real
  propagation environments: Do all antennas contribute equally?'' \emph{IEEE
  Transactions on Communications}, vol.~63, no.~11, pp. 3917--3928, 2015.

\bibitem{mehanna2013joint}
O.~Mehanna, N.~D. Sidiropoulos, and G.~B. Giannakis, ``Joint multicast
  beamforming and antenna selection,'' \emph{IEEE Transactions on Signal
  Processing}, vol.~61, no.~10, pp. 2660--2674, 2013.

\bibitem{konargreed}
A.~Konar and N.~D. Sidiropoulos, ``A simple and effective approach for transmit
  antenna selection in multiuser massive {MIMO} leveraging submodularity,''
  \emph{IEEE Transactions on Signal Processing}, vol.~66, no.~18, pp.
  4869--4883, Sept. 2018.

\bibitem{ibrahim2018learning}
M.~S. Ibrahim, A.~S. Zamzam, X.~Fu, and N.~D. Sidiropoulos, ``Learning-based
  antenna selection for multicasting,'' in \emph{IEEE 19th International
  Workshop on Signal Processing Advances in Wireless Communications (SPAWC)},
  Greece, June 2018, pp. 1--5.

\bibitem{joung2016machine}
J.~Joung, ``Machine learning-based antenna selection in wireless
  communications,'' \emph{IEEE Communications Letters}, vol.~20, no.~11, pp.
  2241--2244, 2016.

\bibitem{park2008capacity}
S.~Y. Park and D.~J. Love, ``Capacity limits of multiple antenna multicasting
  using antenna subset selection,'' \emph{IEEE Transactions on Signal
  Processing}, vol.~56, no.~6, pp. 2524--2534, 2008.

\bibitem{boyd2011distributed}
S.~Boyd, N.~Parikh, E.~Chu, B.~Peleato, J.~Eckstein \emph{et~al.},
  ``Distributed optimization and statistical learning via the alternating
  direction method of multipliers,'' \emph{Foundations and
  Trends{\textregistered} in Machine learning}, vol.~3, no.~1, pp. 1--122,
  2011.

\bibitem{nesterov1983method}
Y.~E. Nesterov, ``A method for solving the convex programming problem with
  convergence rate o (1/k\^{} 2),'' in \emph{Dokl. Akad. Nauk SSSR}, vol. 269,
  1983, pp. 543--547.

\bibitem{francis}
F.~Bach and R.~Jenatton, \emph{Convex Optimization with Sparsity-Inducing
  Norms}.

\bibitem{ibrahim2018prox}
M.~S. Ibrahim, A.~Konar, M.~Hong, and N.~D. Sidiropoulos, ``Mirror-prox {SCA}
  algorithm for multicast beamforming and antenna selection,'' in \emph{IEEE
  19th International Workshop on Signal Processing Advances in Wireless
  Communications (SPAWC)}, Greece, June 2018, pp. 1--5.

\bibitem{demir2014alternating}
{\"O}.~T. Demir and T.~E. Tuncer, ``Alternating maximization algorithm for the
  broadcast beamforming,'' in \emph{Proceedings of the 22nd European, Signal
  Processing Conference (EUSIPCO)}, Portugal, Sept. 2014, pp. 1915--1919.

\bibitem{tran2014conic}
L.-N. Tran, M.~F. Hanif, and M.~Juntti, ``A conic quadratic programming
  approach to physical layer multicasting for large-scale antenna arrays,''
  \emph{IEEE Signal Processing Letters}, vol.~21, no.~1, pp. 114--117, 2014.

\bibitem{bornhorst2011iterative}
N.~Bornhorst and M.~Pesavento, ``An iterative convex approximation approach for
  transmit beamforming in multi-group multicasting,'' in \emph{IEEE 12th
  International Workshop on Signal Processing Advances in Wireless
  Communications (SPAWC)}, CA, USA, June 2011, pp. 426--430.

\bibitem{rusek2013scaling}
F.~Rusek, D.~Persson, B.~K. Lau, E.~G. Larsson, T.~L. Marzetta, O.~Edfors, and
  F.~Tufvesson, ``Scaling up {MIMO}: Opportunities and challenges with very
  large arrays,'' \emph{IEEE Signal Processing Magazine}, vol.~30, no.~1, pp.
  40--60, 2013.

\bibitem{lu2014overview}
L.~Lu, G.~Y. Li, A.~L. Swindlehurst, A.~Ashikhmin, and R.~Zhang, ``An overview
  of massive {MIMO}: Benefits and challenges,'' \emph{IEEE Journal of Selected
  Topics in Signal Processing}, vol.~8, no.~5, pp. 742--758, 2014.

\bibitem{tibshirani1996regression}
R.~Tibshirani, ``Regression shrinkage and selection via the lasso,''
  \emph{Journal of the Royal Statistical Society. Series B (Methodological)},
  pp. 267--288, 1996.

\bibitem{yuan2006model}
M.~Yuan and Y.~Lin, ``Model selection and estimation in regression with grouped
  variables,'' \emph{Journal of the Royal Statistical Society: Series B
  (Statistical Methodology)}, vol.~68, no.~1, pp. 49--67, 2006.

\bibitem{NIPS2008_3418}
F.~R. Bach, ``Exploring large feature spaces with hierarchical multiple kernel
  learning,'' in \emph{Advances in Neural Information Processing Systems 21},
  D.~Koller, D.~Schuurmans, Y.~Bengio, and L.~Bottou, Eds., 2009, pp. 105--112.

\bibitem{jenatton2011structured}
R.~Jenatton, J.-Y. Audibert, and F.~Bach, ``Structured variable selection with
  sparsity-inducing norms,'' \emph{Journal of Machine Learning Research},
  vol.~12, no. Oct, pp. 2777--2824, 2011.

\bibitem{Nemirovski}
A.~Nemirovski, ``Prox-method with rate of convergence {O}(1/t) for variational
  inequalities with lipschitz continuous monotone operators and smooth
  convex-concave saddle point problems,'' \emph{SIAM Journal on Optimization},
  vol.~15, no.~1, pp. 229--251, 2004.

\bibitem{parikh2014proximal}
N.~Parikh, S.~Boyd \emph{et~al.}, ``Proximal algorithms,'' \emph{Foundations
  and Trends{\textregistered} in Optimization}, vol.~1, no.~3, pp. 127--239,
  2014.

\bibitem{boyd_vandenberghe_2004}
S.~Boyd and L.~Vandenberghe, \emph{Convex Optimization}.\hskip 1em plus 0.5em
  minus 0.4em\relax Cambridge University Press, 2004.

\bibitem{Nesterov2005}
Y.~Nesterov, ``Smooth minimization of non-smooth functions,''
  \emph{Mathematical Programming}, vol. 103, no.~1, pp. 127--152, May 2005.

\bibitem{he20121}
B.~He and X.~Yuan, ``On the o(1/n) convergence rate of the douglas--rachford
  alternating direction method,'' \emph{SIAM Journal on Numerical Analysis},
  vol.~50, no.~2, pp. 700--709, 2012.

\bibitem{sion1958general}
M.~Sion \emph{et~al.}, ``On general minimax theorems.'' \emph{Pacific Journal
  of mathematics}, vol.~8, no.~1, pp. 171--176, 1958.

\bibitem{beck2003mirror}
A.~Beck and M.~Teboulle, ``Mirror descent and nonlinear projected subgradient
  methods for convex optimization,'' \emph{Operations Research Letters},
  vol.~31, no.~3, pp. 167--175, 2003.

\bibitem{bubeck2015convex}
S.~Bubeck, ``Convex optimization: Algorithms and complexity,''
  \emph{Foundations and Trends in Machine Learning}, vol.~8, no. 3-4, pp.
  231--357, 2015.

\bibitem{lofberg2004yalmip}
J.~Lofberg, ``Yalmip: A toolbox for modeling and optimization in matlab,'' in
  \emph{IEEE International Symposium on Computer Aided Control Systems Design},
  Taiwan, Sept. 2004, pp. 284--289.

\end{thebibliography}

%

\end{document}